\documentclass[final,authoryear,5p,times,twocolumn]{elsarticle}
\usepackage{graphicx}
\usepackage{amssymb}
\usepackage{amsmath}
\usepackage{natbib}
\usepackage{threeparttable}
\journal{New Astronomy Reviews}

\newcommand\aapr{A\&AR}
\newcommand\aj{AJ}
\newcommand\araa{ARA\&A}
\newcommand\apj{ApJ}
\newcommand\apjl{ApJ}
\newcommand\apjs{ApJS}
\newcommand\aap{A\&A}
\newcommand\pasj{PASJ}
\newcommand\pasa{PASA}
\newcommand\pasp{PASP}
\newcommand\mnras{MNRAS}
\newcommand\nat{Nature}
\newcommand\apss{Ap\&SS}

\newcommand\arcsec{\mbox{$^{\prime\prime}$}}
\newcommand\ion[2]{#1$\;${\scshape{#2}}}%

\def\ergs{erg~s$^{-1}$}
\def\ergcms{erg$~$~cm$^{-2}$~s$^{-1}$}

\begin{document}

\begin{frontmatter}

\title{Ultraluminous X-ray Sources in the {\it Chandra} and {\it XMM-Newton} Era}

\author[ad1]{Hua Feng\corref{cor}}
\ead{hfeng@tsinghua.edu.cn}
\cortext[cor]{Corresponding author}
\address[ad1]{Department of Engineering Physics and Center for Astrophysics, Tsinghua University, Beijing 100084, China}

\author[ad3]{Roberto Soria}
\ead{roberto.soria@mssl.ucl.ac.uk}
\address[ad3]{Curtin Institute of Radio Astronomy, Curtin University, GPO Box U1987, Perth, WA 6845, Australia}

\begin{abstract}
Ultraluminous X-ray sources (ULXs) are accreting black holes that may contain the missing population of intermediate mass black holes or reflect super-Eddington accretion physics. Ten years of {\it Chandra} and {\it XMM-Newton} observations of ULXs, integrated by multiband studies of their counterparts, have produced a wealth of observational data and phenomenological classifications. We review the properties of their host galaxies, list popular spectral models and implications for standard and supercritical accretion physics, demonstrate how X-ray timing of these objects places constraints on their masses. We also review multiwavelength studies of ULXs, including the optical emission of the binary system and nebulosity around them. We summarize that three classes of black holes could power ULXs: normal stellar mass black holes ($\sim 10$~$M_\odot$), massive stellar black holes ($\lesssim 100$~$M_\odot$), and intermediate mass black holes ($10^2$--$10^4$~$M_\odot$). We collect evidence for the presence of these three types of compact objects, including caveat of each interpretation, and briefly review their formation processes.
\end{abstract}


\end{frontmatter}

\section{Introduction}
\label{sec:intro}

\subsection{Early identification of a new class of X-ray sources}

An empirical tenet in the study of accreting compact objects is that their maximum radiative luminosity is about Eddington luminosity, that is the limit where the inward-directed gravitational force is balanced by the outward-directed radiation force. For spherical accretion of fully ionized hydrogen, the Eddington limit can be written as \citep{Frank2002}
\begin{equation}
L_{\rm Edd} =  \frac{4\pi cGMm_{\rm p}}{\sigma_{\rm T}} 
  \approx 1.3 \times 10^{38}\,\left(\frac{M}{M_{\odot}}\right)   
  \rm{erg~s}^{-1} \; ,
\end{equation}
where $\sigma_{\rm T}$ is the Thomson scattering cross section, $m_{\rm p}$ is the proton mass, and $M$ is the black hole (BH) mass.

The X-ray luminosity is generally used as a good proxy for the bolometric luminosity for accreting neutron stars and stellar mass BHs. The Eddington argument proved very valuable in the early stages of X-ray astrophysics, to show that most Galactic\footnote{Throughout this paper, we use the term ``Galactic'' to include also Magellanic Cloud sources.} X-ray binaries are neutron stars \citep{Margon1973}. Later on, dynamical mass measurements confirmed the existence of a population of Galactic BHs with masses $\approx 5$--$15 M_{\odot}$ \citep[][for a review]{Remillard2006}: they also satisfy the Eddington limit, with peak luminosities $\lesssim 10^{39}$ erg s$^{-1}$.

The {\it Einstein} satellite revealed for the first time a small number of non-nuclear X-ray sources in nearby galaxies (particularly, in star-forming galaxies) with apparent luminosities $\gtrsim 10^{39}$~\ergs\ \citep{Long1983,Fabbiano1987,Fabbiano1988,Fabbiano1989,Stocke1991,Stocke1991a}. However, it was not immediately clear whether these sources could be a distinct class from Galactic stellar mass BH candidates, which were just being discovered around the same years.  {\it Einstein}'s relatively crude spatial resolution allowed for the possibility of source confusion in compact star-forming regions; lack of long-term monitoring made it hard to distinguish between persistently luminous sources and transient events, {\it e.g.}, young supernovae (SNe) missed by optical observers.

In the following decade, as nearby galaxies were repeatedly observed with X-ray satellites such as {\it ROSAT} and {\it ASCA}, with superior spectral coverage and resolution, it became clear that some of those non-nuclear sources could not be SNe, and had apparent luminosities well above the Eddington limit of typical Galactic stellar mass BHs. For example, the ``most likely interpretation'' of the brightest non-nuclear source in the spiral galaxy NGC\,1365 was ``an ultra-powerful X-ray binary, with either a highly super-Eddington low-mass black hole or a very massive black hole'' with $M \sim 100$--$200 M_{\odot}$, ``which may pose a challenge for stellar evolution models'' \citep{Komossa1998}; a third scenario, beamed emission, was also proposed for other sources \citep{Okada1998}. Thirteen years later, distinguishing between those three alternative scenarios remains the unsolved fundamental question for this class of sources, despite the wealth of new discoveries from {\it Chandra}, {\it XMM-Newton}, {\it Suzaku} and {\it Swift}, intensive modeling of their X-ray spectral and timing properties, and integrating studies of their multiband counterparts.

For a while, different names have been in use to identify this class of accreting BHs: extraluminous X-ray binaries \citep{Colbert1999}, superluminous X-ray sources \citep{Roberts2000}, intermediate-luminosity X-ray objects \citep{Colbert2002}. In recent years, the community consensus has settled on ultraluminous X-ray sources (ULXs), adopting the terminology first used by Japanese {\it ASCA} teams \citep{Okada1998,Mizuno1999,Makishima2000}.

\subsection{Diversity of the ULX population}

A handful ULXs are observed to exhibit strong X-ray variability at time scales as short as minutes \citep{Strohmayer2003,Strohmayer2007,Heil2009,Feng2010a,Rao2010}, and many others have shown random variation at long time scales from days, weeks to years \citep{Feng2009,Kaaret2009a,Kong2010,Gris'e2010}, confirming their compact nature. They are 10--100 times more luminous than Galactic black hole binaries (BHBs), whose peak luminosity is about $10^{39}$~\ergs. As the Eddington limit is proportional to the BH mass, this suggests that ULXs may contain BHs at least 10--100 times more massive than Galactic ones which are typically of 10~$M_\odot$, and thus represents a new class of BHs with masses in the range of $10^2$--$10^4$~$M_\odot$, intermediate between stellar mass and supermassive BHs. However, some models suggest that the Eddington limit could be violated by a factor up to $\sim$10 for stellar mass BHs \citep{Begelman2002}, and the X-ray luminosity may be overestimated due to anisotropic emission caused by geometric beaming expected at high accretion rate \citep{King2001}. In these circumstances, intermediate mass black holes (IMBHs) are not required to interpret ULXs. A combination of super-Eddington and mildly beamed emission from stellar mass BHs could account for ULXs with an apparent luminosity up to $\sim$$10^{41}$~\ergs\ \citep{Poutanen2007}.

Some objects other than BHBs may appear like ULXs if their timing or multiwavelength properties are unavailable. Young SNe, especially of type IIn, could be as luminous as $10^{40}$~\ergs\ in the X-ray band \citep{Immler2003}. Their X-ray spectrum is usually thermal and soft, and will not show chaotic variability. Young X-ray pulsars could theoretically be more luminous than $10^{39}$~\ergs\ but the number of such sources appears to be very small \citep{Perna2008}. In general, these objects can be distinguished from accreting sources with variability \citep{Kaaret2008}, but it is possible that a SN leaves behind a compact remnant which becomes X-ray luminous when a fallback disk forms \citep{Li2003} and mixed properties may be seen. In a survey of ULXs in nearby galaxies with archival {\it XMM-Newton} data, it was found that the X-ray spectra of a few sources can be decomposed into a power-law component plus a thermal plasma, and in one of them X-ray variability was detected, suggesting that they may be recent SNe with an accreting compact core \citep{Feng2005}. Background active galactic nuclei (AGN) that happen to lie behind a nearby galaxy could be misidentified as ULXs in the galaxy. Their cosmological location could be revealed if redshift is measured from optical lines \citep[\textit{e.g.},][]{Foschini2002a,Masetti2003,Arp2004,Clark2005}. The fraction of background objects is usually high in massive, elliptical galaxies \citep[\textit{e.g.},][]{Tang2005}. Many ULXs are found to be spatially associated with nebulae, young star clusters, or globular clusters; these are most likely located in the galaxy rather than foreground or background objects. Even if they are identified as BHBs, their nature is still not unique in terms of the companion star (low mass or high mass), the mode of the mass transfer (Roche-lobe overflow or wind-fed), or the mass of the accretor (stellar mass or intermediate mass). For example, \citet{Feng2006} found that ULXs in young, star forming galaxies are much more variable than in old, elliptical galaxies (associated with globular clusters). 

No matter what their nature is, ULXs are of great interest in astrophysics. If they are IMBHs, their formation mechanism could be rather different from typical stellar mass BHs; extreme environment, more massive and/or less abundant progenitors are likely required \citep{Madau2001,PortegiesZwart2004}. IMBHs could have played an important role in the formation of supermassive BHs in the early universe and be important targets to search for gravitational wave emission \citep{Ebisuzaki2001,Volonteri2010,Miller2002}. Instead, if they are stellar mass BHs, it is interesting to investigate why they are so distinct from their Galactic cousins. They may be good candidates for the study of the accretion physics at near or super Eddington accretion rates \citep{Ohsuga2005}, and they may shed light on binary stellar evolution channels not seen in our Galaxy.

\subsection{Definition of ULXs}

ULXs are usually defined as non-nuclear, point-like objects which at least once have been observed at an apparent isotropic X-ray luminosity higher than that of stellar mass Galactic BHs: typically, $L_{\rm X} > 10^{39}$~erg~s$^{-1}$ in the 0.3--10~keV band. Some authors \citep[\textit{e.g.}][]{Kaaret2008} prefer to use $L_{\rm X} > 3 \times 10^{39}$ ergs s$^{-1}$, that is the Eddington limit of the heaviest stellar BH ($20 M_{\odot}$) expected at normal metallicity \citep{Belczynski2010}, to exclude a large number of ``faint'' ULXs that should be similar to normal Galactic BHBs.

Such an empirical definition may have ambiguities and also include young X-ray pulsars, SNe, and SN remnants as mentioned above. The consensus in the ULX community is to restrict the definition of ULXs to accreting BHs. Therefore, we define ULXs as {\it non-nuclear accreting BHs with peak luminosities inferred assuming isotropic emission above the Eddington limit of normal stellar mass BHs}. This definition is confined to compact objects that are most likely accreting from a companion star in a binary system, but more exotic possibilities, for example isolated IMBHs accreting via the Bondi process from the surrounding medium are not excluded a priori. Observationaly, variability information is required to define a ULX.

Recently, many ULXs are found to have a mechanical power output as high as in radiation \citep{Pakull2003}. In one of them, the mechanical output is dominant while the radiative output is way below the ULX threshold \citep{Pakull2010}. This source is apparently not a ULX, but speculated to be an offset ULX or quiescent ULX (see Section~\ref{sec:bub} for details). We thus define them as ``ultrapowerful sources (UPSs)'' for completeness, to standard for the same population of sources perhaps in a different state or viewing angle.

\subsection{ULX Catalogs}

A {\it ROSAT}/HRI catalog of luminous, non-nuclear X-ray sources by \citet{Colbert1999} proved to be very important for bringing ULXs to the forefront of research, and for emphasizing the possible observational link with theoretically-predicted intermediate mass BHs, with masses $10^2 \lesssim M \lesssim 10^4 M_{\odot}$. Larger and more systematic X-ray source catalogs based on re-analysis of archival {\it ROSAT}/HRI data soon followed: in particular, by \citet{Roberts2000}, by \citet{Colbert2002}, and by \citet{Liu2005}, which included 106 sources above $10^{39}$ erg s$^{-1}$. The first XMM-Newton catalog was done by \citet{Foschini2002}. A {\it Chandra} catalog with 154 sources in 82 galaxies was produced by \citet{Swartz2004}, and 226 objects were included in the catalog by \citet{Liu2005a}, compiled from mixed literature sources. Such works allowed the first population studies of these objects as a well recognized group, and in relation to the properties of their host galaxies. They also provided the rationale for longer, targeted observations with {\it Chandra} and {\it XMM-Newton}. Today, the largest catalog \citep{Walton2011} is based on the 2XMM Serendipitous Survey, and contains 470 ULXs, of which 367 not listed in previous compilations. Updated catalogs based on {\it Chandra} observations have been compiled by \citet{Swartz2011} and \citet{Liu2011}.

\section{Emission Scenarios and Black Hole Masses}
\label{sec:bh}

\subsection{Different ways of making a ULX}

There are three basic ways to explain non-nuclear systems with apparent accretion luminosities $> 10^{39}$ erg s$^{-1}$, using only standard accretion physics already included in \citet{Shakura1973}'s model. We can either increase the mass of the BH (thus, increasing its Eddington limit), or attribute the enhanced brightness to beamed emission, or require a mass accretion rate high enough to allow the source to exceed the Eddington limit by a factor of a few. It is also possible that the most luminous ULXs may be explained by a combination of all three factors. Assuming solar abundances, the {\it apparent} luminosity\footnote{That is, the observed ${\rm flux} \times 4\pi \times {\rm distance}^2$.} of an accreting BH is:
\begin{align}
&L \approx \frac{1.3 \times 10^{38}}{b} \dot{m} \left(\frac{M}{M_{\odot}}\right) \; \rm{erg~s}^{-1}, &&\dot{m} \lesssim 1 \\
&L \approx \frac{1.3 \times 10^{38}}{b} \left(1+\frac{3}{5} \ln \dot{m}\right) \left(\frac{M}{M_{\odot}}\right)\; \rm{erg~s}^{-1}, \nonumber 
&&1 \lesssim \dot{m} \lesssim 100 \nonumber
\end{align}
\citep{Shakura1973,Poutanen2007}, where $b < 1$ is the beaming factor, and $\dot{m}$ is the dimensionless accretion rate at large radii (typically, the mass transfer rate from the donor star), that is normalized to the Eddington accretion rate\footnote{This definition implicitly includes a ``standard'' radiative efficiency $\eta = 0.1$, typical of efficient disk accretion; hence, a dimensionless accretion rate $\dot{m} \sim 1$ produces a luminosity $L \sim L_{\rm Edd}$. Other authors prefer to define $\dot{m} \equiv \dot{M} c^2/L_{\rm Edd}$, in which case the Eddington limit is reached at $\dot{m} \sim 10$.}: $\dot{m} \equiv \dot{M}/\dot{M}_{\rm Edd} \equiv 0.1 \dot{M} c^2/L_{\rm Edd}$. Let us examine the main tenets of these alternative or complementary scenarios.

{\it Strong beaming} ($1/b \gg 1$) --- A decade ago, accreting BHs with strongly beamed emission were already well known in extragalactic astronomy \citep[BL Lac objects and blazars:][]{Urry1984}, and at least one Galactic BH (SS~433) was known to have semi-relativistic jets, massive outflows and likely anisotropic emission. Naturally, a possible explanation for the early sample of {\it ROSAT} ULXs was that they were microblazars, or SS~433-like objects seen face-on \citep{Fabrika2001,Kording2002,Begelman2006}. The relativistic beaming scenario soon ran into several fatal difficulties. It predicts that for every ULXs, there should also be a large number of lower-luminosity beamed sources: $\approx 30$ sources with apparent $L \sim 10^{39}$ erg s$^{-1}$ for every ULX at $L \sim 10^{40}$ erg s$^{-1}$; but there is no evidence of this population \citep{Davis2004}. Instead, empirical luminosity functions show only $\approx 5$--$10$ sources more luminous than $10^{39}$ erg s$^{-1}$ for every ULX at $10^{40}$ erg s$^{-1}$ \citep{Walton2011,Swartz2004,Grimm2003}, consistent with the higher end of the high-mass X-ray binary luminosity function. There is also no multiband evidence of the large number of jet sources beamed in other directions, required by this scenario \citep{Davis2003}. Conversely, the presence of photoionized bubbles around several ULXs (for example Holmberg II X-1: \citealt{Pakull2002,Kaaret2004,Lehmann2005}; NGC\,5408 X-1: \citealt{Kaaret2009}) requires quasi-isotropic X-ray emission with luminosities $\approx 10^{40}$ erg s$^{-1}$. Finally, the majority of ULXs do not have radio counterparts and/or rapid X-ray variability, expected for relativistic beaming. For all those reasons, relativistic beaming is now ruled out as a general scenario.

{\it Mild beaming and/or super-Eddington accretion} ($1/b \lesssim 10$, $\dot{m} \gg 1$) --- When the accretion rates exceeds the Eddington limit, radiatively-driven outflows are launched from the inner part of the disk, inside the spherization radius where the disk becomes geometrically thick \citep{Shakura1973,Poutanen2007,King2009}. It was suggested \citep{King2001,King2009} that the wind may produce a funnel wall that scatters and collimates the emission along the axis perpendicular to the disk plane. The apparent high luminosity of ULXs is then a combination of this mild collimation (with $b \sim 70/\dot{m}^2$, \citealt{King2009}) and mild super-Eddington emission ($L \sim L_{\rm Edd}(1+3/5\,\ln \dot{m})$). In this scenario, accretion rates $\dot{m} \sim 10$--$30$ are enough to explain most or all ULXs up to luminosities $\approx 10^{41}$ erg s$^{-1}$ with BH masses $\lesssim 20 M_{\odot}$.

Recent numerical radiation-magneto-hydrodynamical simulations \citep{Ohsuga2009,Takeuchi2010,Mineshige2011} supported and advanced earlier analytical results \citep[slim disk models;][]{Abramowicz1988,Watarai2001,Ebisawa2003,Heinzeller2007}, and the radiation-hydrodynamical simulations of, among others, \citet{Ohsuga2005}. An important result of the simulations is that even a moderately super-critical mass supply $\dot{m} \approx 5$ can produce a total luminosity $\approx 1.7 L_{\rm Edd}$ (in rough agreement with the standard analytical scaling used earlier), and an apparent luminosity $\approx 22 L_{\rm Edd}$ for face-on observers, due to mild collimation in the magnetized outflow \citep{Mineshige2011,Ohsuga2011}. 

An alternative explanation for ULXs relies on super-critical accretion but does not require collimation of the emerging radiation. It was suggested that radiation pressure-dominated accretion disks have strong density inhomogeneities due to the development of photon-bubble instabilities \citep{Begelman2002}. Radiation would escape from such disk regions at rates $\sim10$ times above the classical Eddington limit for stellar mass BHs. 

{\it Quasi-isotropic Eddington luminosity} ($1/b \gtrsim 1$, $L_{\rm Edd} \sim 1$) --- Although some models and simulations offer us compelling reasons to ignore the Eddington limit as a constraint for the ULX interpretation, some authors remain wary of this scenario. They point out that the Eddington limit works well for Galactic BHs and neutron stars. More significantly, the isotropic bolometric luminosity of the whole population of $> 60,000$ quasars at redshift $0.2 < z < 4$ in the Sloan Digitized Sky Survey is strongly Eddington-limited at all BH masses, with only a very small fraction of sources between 1 and 3 $L_{\rm Edd}$ \citep{Steinhardt2010}. Perhaps like the bumblebee, the Eddington limit should not fly, but does. 

Thus, it is argued that the null hypothesis for ULXs should be that their luminosities are $\approx L_{\rm Edd}$, despite the difference on the gas reservoir between AGN and binary systems, and their radiative anisotropy is only a factor of $\gtrsim 1$ (a standard accretion disk has a beaming factor $1/b = 2$ in the absence of any other collimation). These conditions are possible to explain most ULXs up to a few $\times 10^{40}$~\ergs\ with accreting BHs of masses $\lesssim 100 M_{\odot}$, which is the theoretical upper limit for a direct collapse of a single stellar core in the current universe (see Section~\ref{sec:bhmass}). Sources more luminous than that may require IMBHs. For example, HLX-1 in ESO243-49, the most luminous ULX found to date, has a peak bolometric luminosity $\approx 10^{42}$ erg s$^{-1}$ \citep{Farrell2009,Davis2011}. If it is Eddington limited, a black hole mass of at least $10^4 M_\odot$ is inferred. Even if super-Eddington emission is allowed, an IMBH is required. 

{\it Quasi-isotropic sub-Eddington luminosity} ($1/b \sim 1$, $\dot{m} < 1$) --- Finally, it is possible that ULXs are accreting at sub-Eddington accretion rates with nearly isotropic emission or disk beaming, which requires the compact objects to be IMBHs ($10^2 M_{\odot} \lesssim M \lesssim 10^4 M_{\odot}$). Then we expect to see similar state transition and X-ray spectral properties from both ULXs and Galactic BHBs, scaled only by the mass.

\subsection{Three types of non-nuclear black holes}
\label{sec:bhmass}

The emission scenarios outlined above require three different ranges of BH masses, with different formation mechanisms.

{\it Ordinary stellar mass BHs} ($M \lesssim 20 M_{\odot}$) --- The first advantage of explaining ULXs with ordinary stellar mass black holes (sMBHs\footnote{To avoid confusion with the acronym "SMBH" usually adopted for supermassive black holes.}) is that we already have direct evidence of their existence in the local universe. There are more than 20 BHs identified in X-ray binaries with dynamically measured masses $\approx 5$--$15 M_{\odot}$ \citep[for reviews see][]{McClintock2006,Remillard2006}. In the Milky Way, the most massive sMBHs are those in GRS~1915$+$105 \citep[$M = 14.0 \pm 4.4 \, M_{\odot}$;][]{Harlaftis2004} and in Cygnus X-1 \citep[$M = 14.8 \pm 1.0 \, M_{\odot}$;][]{Orosz2011}. A similarly massive BH is found in M33 X-7 \citep[$M = 15.65 \pm 1.45$~$M_\odot$;][]{Orosz2007}. An accreting binary system in the nearby starburst galaxy IC 10 \citep{Prestwich2007,Silverman2008} contains a BH with a mass of $\approx 20$--$40 M_{\odot}$ (the mass uncertainty is mainly due to the unconstrained companion mass and inclination angle of the binary plane). The second advantage is that they are the easiest BHs to form: standard stellar models predict sMBHs as the endpoint of individual massive star evolution, for a wide range of stellar masses and metallicities. Thus, the strictest constraints to sMBH models for ULXs come not from the BH formation process, but from the extreme super-Eddington accretion, beaming and outflows required by these models.

{\it Massive stellar BHs} ($20 M_{\odot} \lesssim M \lesssim 100 M_{\odot}$) --- The mass of a compact remnant depends both on the initial mass of the stellar progenitor, and on how much mass is lost through stellar winds during the stellar lifetime \citep{Fryer2001}. Radiatively-driven winds from massive stars are strongly dependent on their metallicity
\citep{Vink2011,Vink2011a}. Therefore, metallicity plays an important role in stellar evolution and determines the maximal mass of a BH that can be formed via core collapse. \citet{Belczynski2010} argued that the collapse of a single star can form a BH up to $\approx 15 M_\odot$ at solar metallicity, $\approx 30 M_\odot$ at 0.3 times solar metallicity, and $\approx 80 M_\odot$ at 0.01 times solar metallicity; consistent results (upper BH mass limit $\approx 70 M_\odot$ at low metallicities) that are obtained by \citet{Heger2003} and \citet{Yungelson2008}. These results can well explain the inferred BH masses in Galactic X-ray binaries (average metallicity close to the solar value) and in the IC 10 source (metallicity $\approx 0.3$ solar). In summary, current stellar evolution models predict a BH remnant distribution peaked around $10 M_{\odot}$, with a tail extending up to $\lesssim 100 M_{\odot}$. As the formation of stellar BHs more massive than $\approx 20 M_\odot$ may require rarer, more extreme conditions, we will refer to them as ``massive stellar black holes (MsBHs)''. 

There is another interesting physical difference between sMBH and MsBH formation, in addition to different progenitor mass and metallicity: sMBHs are probably formed through a SN explosion and subsequent rapid fallback; MsBHs are probably formed from direct core collapse without an explosion \citep{Fryer1999}. This is because for the more massive objects, the gravitational energy released during core collapse is not enough to unbind and expel the stellar envelope. At low metallicities, the threshold for direct collapse is an initial stellar mass $\approx 40 M_{\odot}$ \citep{Fryer1999,Heger2003}; at solar metallicity, even the most massive stars may lose enough mass in winds and explode as SNe. 

{\it Intermediate mass BHs} ($M \sim 10^2$--$10^4$~$M_\odot$) --- To produce an IMBH of $\sim 10^2$--$10^4 M_{\odot}$, the core collapse of an isolated star in the current epoch is not a viable process. This is not because of the lack of very massive stars \citep[in fact, several stars with initial masses $\sim 150$--$300 M_{\odot}$ may exist in the LMC;][]{Crowther2010}, but because Helium cores more massive than $\approx 70 M_{\odot}$ (corresponding to initial stellar masses $\gtrsim 140 M_{\odot}$ at zero metallicities) undergo an electron/positron pair-instability explosion that leads to the complete disruption of the star, without BH formation \citep{Bond1984,Heger2002}. However, a BH may form again if the Helium core is more massive than $\approx 130 M_{\odot}$, corresponding to initial stellar masses $\gtrsim 260 M_{\odot}$ at zero metallicity \citep{Heger2002}.

It has been suggested that metal-free population III stars formed in the very early Universe could reach masses of a few hundred $M_{\odot}$ \citep{Larson1998}, above the pair-instability limit, and thus may have collapsed into IMBHs \citep{Madau2001}. In young and dense star clusters, dynamical friction could lead to massive stars sinking towards the center and undergoing runaway collisions and mergers on timescales $\lesssim 10^6$ yr. Numerical simulations suggest that such a process may result in the formation of a huge star, which may subsequently collapse into an IMBH \citep{PortegiesZwart2002,Gurkan2004,PortegiesZwart2004,Vanbeveren2009}. Dynamical evidence for IMBHs with masses $\sim 10^4 M_{\odot}$ has been proposed for a few globular clusters, \textit{e.g.}, G1 in M31 \citep{Gebhardt2005}, although the issue is still controversial \citep{Anderson2010}. \citet{Vesperini2010} suggested that in massive and/or compact globular clusters, a central seed sMBH may grow by up to a factor of 100 via accretion of gas lost by the first generation of cluster stars in their red-giant phase. IMBHs may also wander in the halo of major galaxies, after tidal stripping of merging satellite dwarfs that contained nuclear BHs \citep{King2005,Bellovary2010}; this is currently the most plausible explanations for the most extreme ULX found to date, HLX-1 in ESO243$-$49.

To summarize, three types of non-nuclear BHs could be powering ULXs (Table~\ref{tab:bh}). 

\begin{table}[h]
\caption{Three types of non-nuclear BHs that could power ULXs by accretion
\label{tab:bh}}
\centering
\begin{tabular}{cclc}
\noalign{\smallskip}\hline\hline\noalign{\smallskip}
Type & Mass & Progenitor & $1/b \cdot L/L_{\rm Edd}$ \\
\noalign{\smallskip}\hline\noalign{\smallskip}
sMBH & $\sim$10 $M_\odot$ & normal star & $\gtrsim 1$ \\
\noalign{\smallskip}\hline\noalign{\smallskip}
MsBH & $\lesssim$100 $M_\odot$ & low-Z star & $\sim 1$ \\
\noalign{\smallskip}\hline\noalign{\smallskip}
 & & pop III & \\
IMBH & $10^2$--$10^4$ $M_\odot$ & cluster core & $<1$ \\
 & & stripped nucleus & \\
\noalign{\smallskip}\hline\noalign{\smallskip}
\end{tabular}
\end{table}

\section{Population studies}
\label{sec:pop}

\subsection{Host galaxy environment}

Assuming that ULXs are binary systems with a BH receiving matter directly from an individual donor star, we can try to fit them into the traditional classification of Galactic X-ray sources, as either high-mass or low-mass X-ray binaries (HMXBs and LMXBs, respectively). HMXBs have a massive OB donor star with a characteristic age $\lesssim 10^7$ yrs; LMXBs have a low-mass donor (often in the subgiant or giant phase) with ages $\gtrsim 10^9$ yrs \citep[\textit{e.g.}][]{Frank2002}. HMXBs are mostly found in spiral and Irr galaxies, with current or recent star formation, while LMXBs populate old spheroidals and globular clusters. ULXs are found both in ellipticals and in spiral/irregular galaxies; however, the most luminous ones are found predominantly in star-forming galaxies.

Specifically, more than two thirds of ULXs found in ellipticals are in the lowest luminosity range, with $L_{\rm X} \lesssim 2 \times 10^{39}$ erg s$^{-1}$ \citep{Swartz2004}, and can easily be explained as the upper end of the LMXB distribution, with BH masses $\sim 20 M_{\odot}$. It was even suggested \citep{Irwin2003,Irwin2004} that all ULX candidates in ellipticals with apparent $L_{\rm X} > 2 \times 10^{39}$ erg s$^{-1}$ could be dismissed as background or foreground sources. Today, it is generally accepted that there are a few genuine ULXs in ellipticals at luminosities $\sim 2$--$10 \times 10^{39}$ erg s$^{-1}$, but none has been found above $10^{40}$ erg s$^{-1}$. In spirals, by comparison, one third of ULXs have luminosities $\gtrsim 4$--$5 \times 10^{39}$ erg s$^{-1}$, and about 10\% have luminosities $\gtrsim 10^{40}$ erg s$^{-1}$ \citep{Swartz2004,Walton2011}. Both early-type (Hubble types 0 to 4) and late-type (Hubble types 5 to 9) spirals may contain ULXs, with similar probability, within the current statistical error of the samples; \citet{Walton2011} find slightly more ULXs in early-type spirals, while \citet{Swartz2011} find slightly more in late types, due to the different spatial resolutions of the instruments.

Furthermore, the fact that a luminous ULX is found in a spiral galaxy does not imply that it has an OB donor star. Many ULXs in spirals are not located in regions (within a few 100s of pc) with evidence of current or very recent star formation \citep{Mushotzky2006}. Using the optical color distribution in a sample of face-on host galaxies, \citet{Swartz2009} found that the brightest ULXs turn on $\sim 10$--$20$ Myr after the end of star formation. Characteristic ages $\gtrsim 10$ Myr for the surrounding stellar population were also found for example in the case of NGC\,1313 X-1 and X-2 \citep{Yang2011,Gris'e2008}, IC 342 X-1 \citep{Feng2008}, NGC\,4559 X-1 \citep{Soria2005}, and NGC 1073 IXO 5 \citep{Kaaret2005}. 

\citet{Swartz2011} normalized the radial distance of all ULX candidates in their sample to the deprojected $D_{25}$ diameter of their host galaxies. They found that the ULX surface density profile is strongly peaked towards the center. The profile can be fitted with a Sersic function, with a best-fit index consistent with a De Vaucouleurs profile ($n = 4$), and an exponential scale-radius $\sim 10^{-3}D_{25}$. This finding suggests that the ULX population is intrinsically related to the stellar population of their host galaxies. Less centrally concentrated distributions would be expected if most ULXs were primordial halo relics such as Population III remnants, or nuclear BHs of accreted satellite galaxies.

A {\it Chandra} sample of ULXs in 58 face-on star-forming galaxies within 15 Mpc, studied in details by \citet{Swartz2009}, does not show statistically significant associations between ULXs and young, massive ($M_{\ast} \gtrsim 10^5 M_{\odot}$) star clusters, within $100\arcsec \times 100\arcsec$ regions. The same conclusion was found by \citet{Zezas2002} for ULXs in the Antennae galaxies. This is inconsistent with formation scenarios based on runaway core collapse and mergers of O stars inside a compact, young stellar cluster, leading to the possible formation of an IMBH of a few hundred solar masses via direct collapse of the resulting superstar \citep{PortegiesZwart2002,PortegiesZwart2004,Freitag2006}. In fact, most ULX host galaxies do not even have stellar clusters sufficiently massive and compact to satisfy the requirements for runaway core collapse. In theory, IMBHs could be formed inside clusters that have since dispersed. However, the evaporation timescale of such clusters would be too long to explain the observed association of many ULXs with young ($\lesssim 20$ Myr) stellar populations.

\subsection{Specific frequency}

Here, we further quantify the association between ULX populations and host galaxy properties. \citet{Swartz2008,Swartz2011} studied 107 ULXs in a complete sample of 127 nearby galaxies ($d < 14.5$ Mpc), selected to be above the completeness limit of both the Uppsala Galaxy Catalog and the {\it Infrared Astronomical Satellite} ({\it IRAS}) survey. The total star formation rate (SFR) of the galaxies in the sample is $\approx 50 M_{\odot}$ yr$^{-1}$ based on the {\it IRAS} data, consistent with other measurements of star formation based on the H$\alpha$ and [\ion{O}{ii}] indicators \citep{Swartz2011}. This, in turn, corresponds to an SFR density $\approx 0.01 M_{\odot}$ yr$^{-1}$ Mpc$^{-3}$ in the local universe. Finding $\approx 110$ ULXs in that volume, after correcting for incompleteness and background interlopers, implies that we expect, on average, $\approx 2$ ULXs for a SFR of $1 M_{\odot}$ yr$^{-1}$, and 1 ULX in a volume of $\approx 50$ Mpc$^{-3}$. This is consistent with the HMXB luminosity function in local star-forming galaxies: \citet{Grimm2003} find $\approx 30$ sources with $L_{2-10} \gtrsim 5.5 \times 10^{38}$ erg s$^{-1}$ (corresponding to $L_{0.3-10} \gtrsim 10^{39}$ erg s$^{-1}$ in the ULX definition used in \citealt{Swartz2011}) in a sample of galaxies with a total SFR $\approx 16 M_{\odot}$ yr$^{-1}$.

The average specific frequency in star-forming galaxies, in the local universe, is 1 ULX per $\approx 10^{10} M_{\odot}$, for galaxies with dynamical mass of $10^{10} M_{\odot}$ \citep{Swartz2008,Swartz2011,Walton2011}. This rate decreases for increasing galaxy mass, scaling as $M^{-0.6}$ \citep{Swartz2008,Walton2011}. In elliptical galaxies, the specific rate is 1 ULX per $\approx 10^{11} M_{\odot}$ \citep{Walton2011}. Thus, it appears that low-mass spirals and irregulars are more efficient at producing ULXs, per unit mass. This is true at least down to galaxy masses $\sim 10^{8.5} M_{\odot}$. 

There may be two reasons why low-mass galaxies contain more ULXs per unit mass \citep{Walton2011}. One is that their SFR per unit mass is higher, so they are more efficient at producing ULXs. The other is that they have lower metal abundances, which may favor the formation of heavier stellar BHs from the collapse of massive O stars \citep{Zampieri2009,Mapelli2009,Soria2005,Pakull2002}. However, it is suggested by \citet{Prestwich2010} that the metallicity does not affect the ULX production strongly after comparing two pairs of colliding galaxies.

Finally, it is often noted that many luminous ULXs reside in colliding or strongly interacting galaxies: the Antennae \citep{Zezas2002a,Zezas2002}, the Cartwheel \citep{Gao2003,Wolter2004,Wolter2006,Crivellari2009}, the Mice \citep{Read2003}, NGC\,4485/90 \citep{Roberts2002}, NGC\,7714/15 \citep{Soria2004a,Smith2005}, NGC\,3256 \citep{Lira2002}, and a few more Arp galaxies. This may simply be due to the fact that colliding galaxies have the largest SFR, and the number of ULXs scales with the recent SFR. 

\subsection{Luminosity functions}

The cumulative luminosity function (LF) of a population of accreting sources (also known as $\log N(>S)$--$\log S$ curve) is expected to have a break or cut-off at the Eddington luminosity of the most massive objects in the population. Therefore, in the absence of direct mass measurements, the LF can be used as a rough indicator of the maximum mass. In fact, things are not so simple. The LF of X-ray binaries in our Galaxy and in other nearby galaxies does not show any features at $\approx 2 \times 10^{38}$ erg s$^{-1}$, Eddington limit of the neutron star population. Moreover, the location of the upper cutoff may be a function both of mass (via the Eddington limit) and of the age of the population, since the most luminous sources tend to have a shorter lifetime \citep{Wu2001,Kilgard2002}. Thus, careful modeling is required to interpret a LF.

\begin{figure}[t]
\centering
\includegraphics[height=\columnwidth,angle=270]{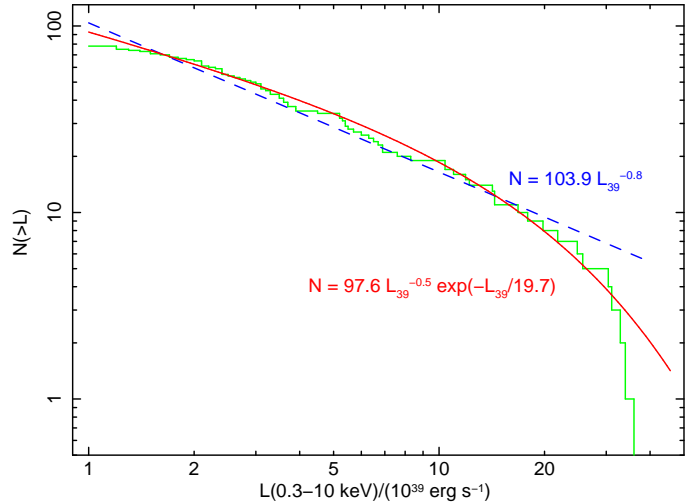}
\caption{Cumulative ULX luminosity function from the {\it Chandra} survey of \citet{Swartz2011}. For this sample, a cut-off power-law with a break $\approx 2 \times 10^{40}$~\ergs\ matches the observed distribution better than an unbroken power-law, although the issue remains controversial \citep[\textit{cf.}][]{Walton2011}. At the low-energy end, it also matches the slope and normalization of the HMXB distribution below $10^{39}$\ergs\ \citep{Grimm2003}.
\label{fig:lf}}
\end{figure}

With this caveat, we can use the ULX LF to test the scenario that ULXs in star-forming galaxies are mostly the high-luminosity tail of HMXBs (plus some additional LMXBs), and ULXs in ellipticals are the high-luminosity tail of LMXBs. Several {\it Chandra} and {\it XMM-Newton} surveys of discrete X-ray sources in nearby galaxies have produced consistent results in agreement with this interpretation \citep{Grimm2003,Colbert2004,Swartz2004,Walton2011}. In summary, they have all found that the cumulative LF of young X-ray populations in star-forming galaxies is a simple power-law with a slope $\approx 0.6$--$0.8$, while the cumulative LF of old X-ray populations in ellipticals and spiral bulges has a slope of $\sim 1.5$. Thus, short-lived HMXB-like sources dominate the high-luminosity end of the distribution. The normalization of the HMXB LF is proportional to the SFR; the normalization of the LMXB LF is proportional to the stellar mass \citep{Gilfanov2004a,Gilfanov2004}.

What remains controversial is the behavior of the LF for the young ULX population above an X-ray luminosity $\approx 10^{40}$ erg s$^{-1}$. From their {\it Chandra} survey, \citet{Swartz2011} find that a pure power-law $N(>L_{\rm X}) \sim L_{\rm X}^{-\alpha}$ provides a very poor fit to the data above that threshold (Figure~\ref{fig:lf}). A cut-off power-law model with $N(>L_{\rm X}) \sim L_{\rm X}^{-\alpha} \exp(-L_{\rm X}/L_{\rm c})$ provides an acceptable fit, with $\alpha \approx 0.5$ (as required by the low-luminosity end of the LF) and $L_{\rm c} \approx 2 \times 10^{40}$ erg s$^{-1}$. A very similar result was also found by \citet{Grimm2003}, \citet{Swartz2004}, and \citet{Mineo2011}. The cut-off power-law model does not preclude the possibility of finding ULXs above $2 \times 10^{40}$ erg s$^{-1}$, but they would be increasingly rarer. It predicts \citep{Swartz2011} 1 ULX with $L_{\rm X} \gtrsim 5 \times 10^{40}$ erg s$^{-1}$ within a radius of $\sim 13$--$29$ Mpc, and 1 ULX with $L_{\rm X} \gtrsim 10^{41}$ erg s$^{-1}$ within a radius of $\sim 42$--$119$ Mpc. If the luminosity of a few ULXs recently discovered by \citet{Sutton2011} is confirmed to be $\approx 10^{41}$ erg s$^{-1}$, their existence would not be consistent with the cut-off model of \citet{Swartz2011}, and may suggest a physically different population of accretors \citep{King2005}. And, of course, so does HLX-1 in ESO243$-$49 with its peak outburst luminosity $\approx 10^{42}$ erg s$^{-1}$ \citep{Farrell2009,Davis2011}. 

In contrast with those findings, the {\it XMM-Newton} survey of \citet{Walton2011} does not see any turnover or cut-off in the cumulative LF above $10^{40}$ erg s$^{-1}$, but their data cannot rule out the presence of a similar break. At this stage, the discrepant result may be caused by a different binning technique, or by small-number statistics. Both the \citet{Walton2011} and \citet{Swartz2011} complete subsamples have only 5 sources with $L_{\rm X} \gtrsim 2 \times 10^{40}$ erg s$^{-1}$ out of a total of $\approx 100$ ULXs. A few other ULXs above that threshold are known, but cannot simply be added to the LF because they do not form part of a complete sample.

\section{X-ray Spectral Properties}
\label{sec:spec}

Thanks to the large effective area of {\it XMM-Newton}, combined with {\it Chandra} observations, the spectral and timing properties of ULXs in the $0.3$--$10$ keV band have been widely investigated and modeled over the last decade, even though the physical interpretation is still controversial. Most of the X-ray studies have focused on bright ULXs ($f_{\rm X} \gtrsim 10^{-12}$~\ergcms) in nearby galaxies ($\lesssim 10$~Mpc) that allows spectral and timing analysis at high signal to noise ratios. In the next two sections, we review X-ray spectral and timing properties of bright ULXs  and discuss their physical interpretation in comparisons with well-defined emission properties of Galactic BHBs.

\begin{figure}[t]
\centering
\includegraphics[width=\columnwidth]{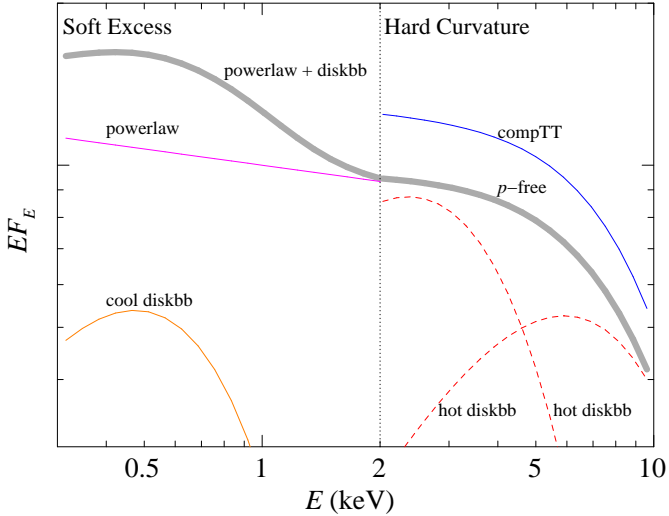}
\caption{A typical ULX spectral shape (grey) in 0.3-10 keV, with a soft excess below 2 keV and a hard curvature above 2 keV. The soft excess can be modeled by a cool thermal component, over the power-law extension of the hard component. A slim disk model ($p$-free) or a warm, thick Comptonization model can adequately fit the hard curvature; in the plot, a $p$-free model with $kT_{\rm in} = 2.5$~keV and $p = 0.5$ has an almost identical spectral shape to a compTT model with $kT_{\rm e} = 2.04$~keV and $\tau = 6.39$, which is shifted vertically for clarity. For comparison, two diskbb models are plotted with $kT_{\rm in} = 1.0$ and 2.5~keV, respectively. They are more curved and cannot fit the energy spectrum over the 0.3-10 keV band.
\label{fig:spec}}
\end{figure}

Phenomenologically, the energy spectra of ULXs in the $0.3$--$10$ keV band can be divided into two groups: those that are consistent with a simple power-law, and those that are more complex \citep{Makishima2007}. The latter mainly refers to a mild broad curvature (convex shape) over the whole band, a break or steepening above $\sim 2$ keV, or a soft excess below $\sim 2$ keV. A typical ULX spectral shape of the latter is sketched in Figure~\ref{fig:spec}.

\subsection{Power-law spectra: hard or steep power-law state}

Many ULX spectra are well fitted by a single, absorbed power-law model. The photon index has a broad distribution, peaked at $\Gamma \approx 1.8$--2.0 \citep{Swartz2004,Winter2006,Berghea2008} but with a few much harder (down to $\Gamma \approx 1$) as well as much softer sources (up to $\Gamma \approx 3$). The distribution does not appear to be bimodal; there is no evidence of a gap between a ``hard state'' and a ``soft state''. \citet{Berghea2008} found that the power-law photon index tends to be harder at increasing luminosity; a selection of ULXs with high luminosity ($L_{\rm X} > 10^{40}$ erg s$^{-1}$) and hard power-law spectrum ($\Gamma \lesssim 1.8$) is also presented in \citet{Soria2011}.

Some of the hard sources with multiple observations show strong flux variability (sometimes by an order of magnitude) but constant photon index \citep{Feng2009,Kaaret2009a,Soria2009}. Similar behavior has been seen from Galactic BHBs in the hard state \citep{Belloni2005,Remillard2006}. The X-ray luminosity in the hard state is usually $\lesssim 0.03$ times Eddington \citep{Fender2004}, but during hard-to-soft state transitions the hard state can reach a luminosity up to 30\% of Eddington \citep{Miyakawa2008,Yu2009}. Thus, if hard power-law ULXs belong to the same canonical hard state defined in Galactic BHs, their BH masses must be $\gtrsim 10^3 M_{\odot}$ \citep{Winter2006}.

A key signature of the hard state is the presence of compact, continuous radio jets. So far, the only possible detection of such jets is from IC 342 X-1, with {\it VLA} observations \citep{Cseh2011}. Using the fundamental plane of accreting BHs, \citet{Cseh2011} estimate a compact object mass $(1.2$--$13.6) \times 10^3 M_{\odot}$. Further observations are needed to spectrally constrain the nature of the radio core in this system.

M82 X-1 is another good candidate for hard state ULXs. This source displays low frequency quasi-periodic oscillations (QPOs) \citep{Strohmayer2003} above a band-limited broad noise component, and its X-ray spectrum is a featureless power-law with a photon index $\approx 1.7$ \citep{Kaaret2006}. We note that M82 X-1 may be one of the very few ULXs that change their spectral state during outbursts, switching from a hard to a thermal state (see Section \ref{sec:curve}).

The hard spectrum may come from an advection-dominated accretion flow \citep{Esin1997}, a luminous hot accretion flow \citep{Yuan2001}, a classical hot corona above the accretion disk \citep{Liang1977}, or from a combination of synchrotron and inverse-Compton emission at the base of the jet \citep{Markoff2001}. The first mechanism is limited to low accretion rates, but the latters can in principle operate also near the Eddington luminosity and be relevant to ULXs \citep{Freeland2006}. Thus, an alternative interpretation favored by some authors is that hard power-law ULXs are not in the canonical low/hard state; in other words, that the ULX hard state is not confined to luminosities $\lesssim 0.1$ Eddington, and does not switch to a softer, disk-dominated thermal state as the accretion rate increases \citep{Soria2011}. We could call it a persistent high/hard state (unknown in Galactic BHBs), which can extend all the way to the Eddington luminosity and smoothly morph into the very high state or other super-Eddington accretion states.

At the opposite end of the broad photon-index distribution, several ULXs with $\Gamma \gtrsim 2.5$ could be classified in the steep power-law state of Galactic BHBs, typically found at near-Eddington luminosities \citep{Feng2005,Winter2006,Soria2007}. The fact that there are no clear gaps, but instead a continuous distribution in photon index and an overlapping luminosity distribution between hard and soft ULXs may be an indication that the two regimes are smoothly connected.

In conclusion, with longer X-ray observations and increasing photon statistics, we have reached a stage where we can identify and model a number of subtle spectral features on top of the simple power-law for many ULXs. In the rest of this section, we discuss a variety of spectral models that account for such features, and their physical implications. 

\subsection{Soft excesses: cool disk emission or massive outflow?}

In many ULXs, excessive residuals at low energies ($<2$~keV) would be seen if one makes a power-law fit at high energies; an additional disk component could adequately fit the excesses and the inferred disk inner temperature is around 0.1--0.4~keV. This cool, soft excess was first identified by \citet{Fabian1993} in NGC 5408 X-1; \citet{Kaaret2003} showed the emission was point like and applied a disk plus power-law model. Similar phenomena were then found in NGC 1313 X-1 and X-2 \citep{Miller2003}, and later on in many other ULXs \citep{Miller2004,Feng2005,Stobbart2006,Winter2006}. Such a fit is analogous to the empirical model for Galactic BHBs except that the disk is cooler and more luminous in ULXs than in sMBHs at their high state \citep{Gierli'nski2004}. For a standard accretion disk \citep{Shakura1973} extending to the last stable orbit around a BH, the inner disk temperature $T_{\rm in}$ is a function of BH mass $M$ and total luminosity $L$:
\begin{equation}
  T_{\rm in} = 1.2 \left(\frac{\xi}{0.41}\right)^{1/2}
  \left(\frac{\kappa}{1.7}\right) \alpha^{-1/2}
  \left(\frac{L}{L_{\rm Edd}}\right)^{1/4}
  \left(\frac{M}{10 M_\odot}\right)^{-1/4} \, {\rm keV,}
\end{equation}
where $\alpha$ depends on the BH spin ($\alpha =1$ for a Schwarzschild BH, $\alpha = 1/6$ for an extreme Kerr BH), $\xi$ is a corrective factor that takes into account the no-torque condition at the innermost orbit, and $\kappa$ is the spectral hardening factor \citep{Makishima2000}. Given the same spinning and correction factors, the BH mass is scaled with the disk inner temperature and luminosity as a function of
\begin{equation}
  M \propto T_{\rm in}^{-2} L^{1/2}
\end{equation}
Thus, the low-temperature, high-luminosity characters of the assumed disk suggest that the emission arises from accretion onto IMBHs. For comparison, the typical disk inner temperature is about 1~keV at a luminosity of $10^{38}$~\ergs\ for sMBHs about 10~$M_\odot$. The characteristic temperature for ULXs is usually around 0.1--0.4~keV and the luminosity is some $10^{39}$--$10^{40}$~\ergs, suggestive of a BH mass of $10^3$-- $10^4$~$M_\odot$.

However, the interpretation of soft excesses as emission from cool accretion disk has confronted challenges. \citet{Gonccalves2006} argued that soft excesses could also be soft deficits depending on the energy band where the power-law is fitted. \citet{Berghea2008} found similar soft excesses in both ULXs and less luminous X-ray sources, implying that they are not a unique signature of IMBHs in a statistical sense. It is suggested that supercritical accretion may produce massive outflows, which are optically thick with soft, thermal emission at similar temperature, being another possible origin for the detected soft excesses \citep{King2004,Begelman2006,Poutanen2007}. Other authors \citep{Roberts2007,Soria2007,Gladstone2009,Weng2011} have suggested that the soft excess does come from the disk surface, but much further away than the innermost stable orbit, where the disk may be obscured or replaced by an outflow or a scattering corona (Section~\ref{sec:curve}). In that case, the characteristic radius of the thermally-emitting disk region cannot be used to infer the BH mass.

Repeated observations of an individual ULX may help test the cool disk model \citep{Soria2007,Feng2007a}. If the disk touches the innermost stable orbit around the BH, one expects to see the disk luminosity vary to the 4th power of the inner temperature, $L_{\rm disk} \propto T_{\rm in}^4$, as well known in the thermal state of Galactic BHBs \citep{Gierli'nski2004}. If, instead, the observed inner disk radius is the boundary between an outer standard disk and an inner corona or outflow, we expect that such transition will move further and further out at higher accretion rates when super-Eddington. In the simplest outflow scenario, $R_{\rm in} \sim \dot{m}$ \citep{Begelman2006,Poutanen2007}. This implies \citep{Soria2007} $L_{\rm disk} \approx$ constant for a viscously heated outer disk and $L_{\rm disk} \propto T_{\rm in}^{-4}$ for a strongly X-ray-irradiated outer disk (as is more likely the case in ULXs). Most other plausible relations between $R_{\rm in}$ and $\dot{m}$ predict negative correlations between the disk luminosity and temperature, with slopes between $0$ and $-4$ \citep{Poutanen2007}. A negative correlation between the luminosity and temperature is clearly detected for at least NGC 1313 X-2 and IC 342 X-1 \citep{Feng2007a,Feng2009}. From a sample of 8 ULXs, \citet{Kajava2009} found that the luminosity of the soft excesses scales with the temperature as $L_{\rm soft} \propto T^{-3.5}$, roughly consistent with the model prediction that the soft emission arises from a photosphere due to strong outflows from a supercritical disk and its temperature is determined by the spherization radius \citep{Poutanen2007,Soria2007}.  On the contrary, a positive correlation is seen from NGC\,5204 X-1; however, it has a shallower slope $L \propto T_{\rm in}^{2.1 \pm 0.5}$ \citep{Feng2009}. \citet{McClintock2009} suggested, based on a disk atmosphere model, that the hardening correction factor becomes larger at increasing luminosity, and explains the apparent shallower slope in Galactic BHs such as XTE~J1550$-$564 and H1743$-$322. Applying such a varying hardening correction, one can recover the $L \propto T^4$ relation for NGC\,5204 X-1, which would favor the IMBH scenario. However, this result is not robust yet: \citet{Kajava2009} found that the positive correlation disappears or even becomes negative if one chooses a different lower energy bound for the spectral fitting. Also, the disk contributes only $< 25\%$ of the 0.3--10 keV luminosity, inconsistent with the thermal dominant state. 

Instead, if ULXs were IMBHs, we would expect to see many of them in the canonical thermal state dominated by the soft disk component. In summary, using the disk parameters inferred from X-ray spectra to determine the BH mass is based on assumptions that have proved either incorrect or problematic for most sources. 

\subsection{High energy curvature: hot standard disk, slim disk, or warm corona?}
\label{sec:curve}

It was noted by \citet{Stobbart2006} and \citet{Gladstone2009} that high energy curvature/steepening is a common feature in high-quality ULX spectra. Generally, the break/steepening could be fitted by a Comptonization model with relatively low energy electrons and high optical depth \citep{Makishima2007,Roberts2007}. A few ULX spectra are mildly convex, formally consistent with a standard disk model. We thus discuss the physical implications of these models.

\subsubsection{Hot standard disk and thermal state}
With {\it ASCA} observations, a hot multicolor accretion disk model was found to fit the spectra of a number of ULXs \citep[\textit{e.g.}][]{Kubota2001}. For those with relatively low temperatures ($kT_{\rm in} \lesssim 1$~keV), they could be explained by accretion onto sMBHs or MsBHs \citep{Winter2006}. The high temperature and luminosity may be explained by large BH spin \citep{Makishima2000}, but this does not work for extreme cases where relatively small BH mass is required, which in turn is inconsistent with the observed high luminosity.  For {\it XMM-Newton} spectra, a single hot disk with absorption is usually unable to fit the data; an additional component is required and dominates in the low energy band \citep{Roberts2005,Feng2005,Stobbart2006,Winter2006}, and the disk component is not the dominant part \citep{Stobbart2006}. The presence of a soft component in addition to the hot disk is unusual in BHBs and does not have a simple physical interpretation. In general, the hot disk model is problematic.

The standard disk emission is best known in the thermal state in BHBs, in which the X-ray spectrum is dominated by emission from a standard accretion disk; plus, the X-ray variability is low, with no or weak QPOs \citep{Remillard2006}. 
So far, the thermal state may have been identified only in one case, M82 X-1, constrained using both spectral and timing information. With simultaneous {\it Chandra} and {\it XMM-Newton} observations of M82, \citet{Feng2010} found surprisingly that the previously known QPOs in the source disappeared. The light curve was no longer highly variable and the power spectrum was consistent with that of white noise. The energy spectrum also changed dramatically from a straight power-law to a disk-like spectrum. Fitting with a standard accretion disk model, the inferred luminosity and temperature from three repeated observations are consistent with a 4th power relation. All results are well consistent with that expected for the thermal state. The monitoring data from {\it RXTE} indicate that these {\it Chandra} and {\it XMM-Newton} observations were made during the source outbursts, suggesting that M82 X-1 usually stays in the hard state and could transition to the thermal state during outbursts. Fitting with a relativistic disk model, the BH is estimated to be a fast spinning IMBH of 200--800~$M_\odot$.

There are a few more sources \citep{Isobe2008,Kajava2009,Jin2010} which have shown spectral variability possibly consistent with a $L \propto T^4$ relation if fitted with a hot standard disk model. Another group of five nearby ULXs consistent with the thermal state spectrum, based on X-ray spectral modeling, was discussed by \citet{Hui2008}. These sources all lie at the lower end of the ULX luminosity regime, and could be explained by Kerr MsBHs in the thermal state. They do not stand for the majority of ULXs, in particular those with a typical luminosity around $10^{40}$~\ergs. Also, lack of fast timing information hampers a firm identification.

Therefore, it is generally accepted that the canonical thermal state is rare in ULXs \citep{Soria2009a}.

\subsubsection{Slim disk and $p$-free model}

The slim disk can radiate near or moderately above the Eddington limit estimated from spherical accretion \citep{Abramowicz1988}. This super-Eddington disk is thus a good candidate for ULX emission \citep{Watarai2001,Ebisawa2003}. When advective energy transport dominates over radiative cooling in an accretion disk, the radial temperature profile changes from $T \propto R^{-3/4}$ for a standard disk to $T \propto R^{-1/2}$ for a slim disk \citep{Watarai2000}. An extended multicolor disk model called the $p$-free model was introduced, with the radial dependence $p$ on temperature ($T \propto R^{-p}$) being an additional free parameter \citep{Mineshige1994}, and used to explain ULX emission \citep{Watarai2001}. \citet{Vierdayanti2006} investigated the energy spectra of four ULXs, NGC 5204 X-1, NGC 4559 X-7/X-10, and NGC 1313 X-2, and found that they can be well fitted by the $p$-free model. Interestingly, the best-fit $p$ value was found to be around 0.5, consistent with that for a slim disk. In the luminosity vs.\ disk temperature diagram, the four ULXs and Galactic BHBs appear to be in line with the same $L \propto T^4$ relation, suggesting that ULXs may be BHs of similar or slightly higher masses (sMBHs or MsBHs) but in the slim disk state. Another successful test of the $p$-free or slim disk model was for M33 X-8, which is marginally classified as a ULX but a good candidate for deep investigation owing to its relatively small distance \citep[\textit{e.g.,}][]{Foschini2004,Foschini2006}. \citet{Weng2009} first identified that an extra power-law component is required in addition to the $p$-free disk to fit the source spectrum. The best-fit $p$ value ($0.57 \pm 0.03$) suggests notable advection in the disk, close to the slim disk nature rather than a standard one. These indicate that the slim disk state may be a natural extension of the canonical thermal state at supercritical accretion. This is consistent with the radiation-magnetohydrodynamic simulations that the hard, thermal and slim states could be produced under a single model with different accretion rates \citep{Ohsuga2009}.

However, the $p$-free model encounters difficulties in both observation and theory.  In some sources, the disk inner temperature was found too high to be physically reasonable when fitted with the $p$-free model \citep{Gladstone2009}. If photon trapping is taken into account, the radial temperature profile is expected to be significantly flatter than $p=0.5$ within the trapping radius \citep{Ohsuga2002,Ohsuga2005}, in contrast to the spectrum predicted without considering photon trapping \citep{Watarai2000}. Also, numerical simulations suggest that dense, warm outflows are always associated with supercritical accretion \citep{Ohsuga2005,Ohsuga2009}, which would Comptonize and obscure the emission from the underlying disk, see details in the following subsection.

\subsubsection{Warm corona and ultraluminous state}

In Galactic BHBs, high energy spectral curvature is also detected, but usually occurs at tens to hundreds of keV. It is interpreted as Comptonization in a thin corona with temperatures typical of $10^2$~keV. However, the spectral curvature in ULXs suggests an electron temperature of a few keV and a scattering optical depth significantly higher than unity, indicative of a warm and thick corona. Due to the presence of such a thick corona, the emission from the disk would be distorted and the previous interpretation, a simple summation of a cool disk and a warm corona \citep{Stobbart2006}, needs be modified consequently.

\citet{Gladstone2009} revisited best quality {\it XMM-Newton} data and concluded that two common features simultaneously exist in the energy spectrum of ULXs: soft excesses below 2 keV and high energy curvature above 2 keV. This is not frequently seen in the spectra of Galactic BHBs when they are thought to accrete at sub-Eddington level. Thus, they suggested that ULXs represent a new state distinct from the four well-known states (quiescent, hard, thermal, and steep power-law) in Galactic BHBs \citep{Remillard2006}, and call it ultraluminous state. Please refer to \citet{Roberts2010} for a more detailed review of this state.  The Galactic BHB that comes closest to this state is XTE~J1550$-$564 at the peak of its 1998 September outburst. A coupled disk-corona model \citep{Svensson1994,Done2006} was proposed to explain such state, in which the disk and corona are not independent but share the total gravitational energy release. The un-scattered outer disk is responsible for the soft excesses, while the Comptonized inner disk produces the hard, curved tail. The BH mass, estimated from the unmasked temperature and radius of the underlying disk, lies in the regime of sMBHs or MsBHs. 

Numerical simulations of supercritical accretion indicates that radiation pressure would power strong outflows above the disk in the inner region \citep{Ohsuga2005,Ohsuga2009}. The wind is optically thick and warm (a few keV), and would strongly Comptonize the disk emission. At high accretion rate, the disk becomes thick and would collimate the radiation along the symmetric axis causing geometric, mild beaming with an apparent isotropic luminosity over the Eddington limit by a factor up to $\sim$10. This picture is consistent with the proposed ultraluminous state. Observationally, from the spectral evolution of M33 X-8, \citet{Middleton2011a} suggested that the outflow is launched at near-Eddington luminosity, and the radius at which it is launched moves further out with increasing luminosity.

\subsubsection{Bulk motion Comptonization}

If the Comptonizing region is the fast outflow, rather than a static corona, it is possible that bulk motion \citep{Titarchuk1998} can be as or more important than thermal Comptonization, depending on the characteristics of the outflow \citep{Kawashima2009}. Observationally, ULX spectra that can be formally fitted with thermal Comptonization models can be equally well fitted with bulk motion models \citep[\textit{e.g.}][]{Roberts2005}. The only difference is in the interpretation of the characteristic energy of the upscattering electrons. As noted by \citet{Soria2011}, simple physical considerations, comparisons with other astrophysical classes of outflows, and numerical simulations all suggest that super-Eddington ULX outflows have a characteristic speed $\sim 0.1c \sim$ escape velocity from the launching radius, corresponding to a bulk kinetic energy $\sim 5$ keV for the outflowing electrons. This is similar to the characteristic energy of the spectral breaks. 

\subsection{Supersoft ULXs}

\begin{figure}[t]
\centering
\includegraphics[width=\columnwidth]{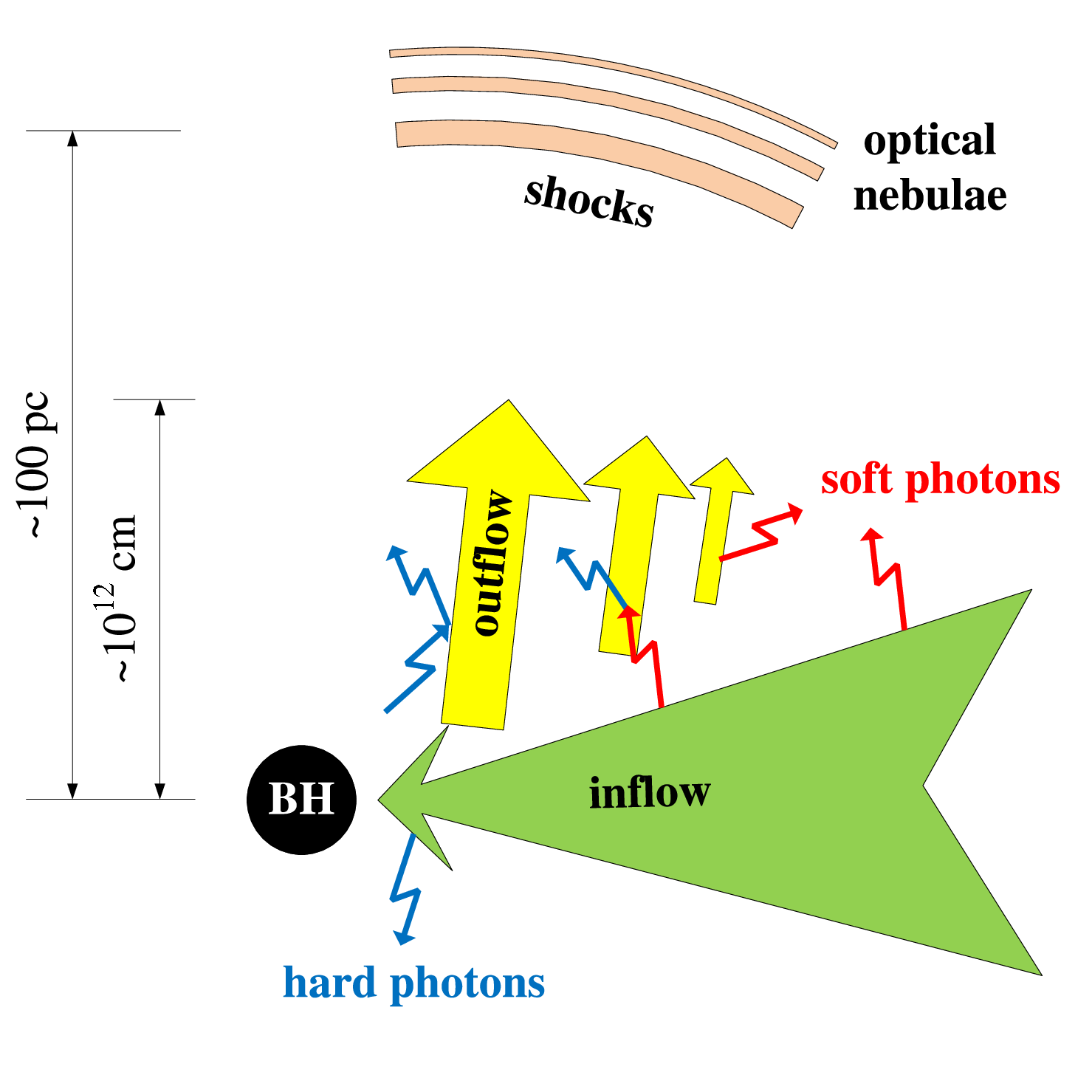}
\caption{A toy model of ULX. The source is powered by supercritical accretion onto a black hole. Radiation pressure propels strong outflows around the inner region of the accretion flow, where a slim disk may exist in connection with a standard disk far outside. The outflow is warm (a few keV) and optical-thick and may scatter off and redirect photons, froming a funnel with geometric beaming; it could also up-Comptonize soft photons; the inner disk and/or the outflow are responsible for the observed hard emission component above 2 keV with a curvature. The soft excess may arise from the outer disk or the optical-thick outflow. If viewing along the disk plane, one may only see the soft emission of the outflow and this could be one scenario for the supersoft ULXs. When the outflow hits the interstellar medium, it may produce shocks seen as optical nebulae. The picture is not to scale.
\label{fig:model}}
\end{figure}

Most ULXs and BHBs have considerable radiation output above 2~keV, while supersoft sources (SSSs) have most of their energy released below 2~keV. Canonical SSSs are speculated to be powered by steady nuclear burning on the surface of white dwarfs \citep{DiStefano2010}. A number of supersoft ULXs have been discovered, whose spectrum is dominated by a cool blackbody component of tens to a hundred eV, being a distinct class from the majority of ULXs. These include M101 ULX-1 \citep{Pence2001,Kong2005,Mukai2005}, Antennae X-13 \citep{Fabbiano2003}, M81 ULS1 \citep{Swartz2002,Liu2008}, NGC 4631 X1 \citep{Carpano2007,Soria2009}, and NGC 247 ULX \citep{Jin2011}.  At long timescales, most of these sources have shown dramatic variability with a luminosity change by a factor up to $10^3$ but with consistently low temperatures. A weak power-law tail is significantly detected in at least one source \citep{Jin2011}, but none of the supersoft ULXs has become a power-law dominated ULX or vice versa. Fast timing noise was detected in two of them, M101 ULX-1 \citep{Mukai2005} and NGC 247 ULX \citep{Jin2011}, down to a timescale around 100~s, indicating that they are not AGN. The nature of these peculiar sources is very uncertain. They are unlikely powered by white dwarfs, because they are too luminous (a few times $10^{39}$~\ergs) and some of them are too hot ($\gtrsim 100$~eV). They are also unlikely to be due to cool disks around IMBHs, because such disks cannot have a huge variability with a constant temperature. One possible scenario is that their emission is produced by massive outflows from supercritical accreting sources viewed at a high inclination angle \citep{Ohsuga2005,Poutanen2007}. A unification model (supercritical accretion with thick outflows, see Figure~\ref{fig:model}) may be able to explain both normal, broad-band ULXs and less luminous, supersoft ULXs with different viewing angles.

So far, we have briefly reviewed a variety of scenarios that can account for the observed ULX spectra (see also Figure~\ref{fig:spec}). In Figure~\ref{fig:model}, we sketch the physical ingredients behind the most common combination of such models, which may be successfully used to explain the multiband spectrum of most ULXs.

\subsection{State transitions of ULXs}

The study of long term spectral variability of ULXs mainly relies on pointed observations with {\it Chandra} or {\it XMM-Newton}, or quasi-regular snapshots with the X-Ray Telescope on board {\it Swift}. In addition, {\it RXTE} has been used to monitor the brightest (in terms of observed flux) ULX in M82. The most interesting science with these observations, in addition to the search for orbital periods, is perhaps to find state transitions known in Galactic BHBs. 

Canonical transitions between the low/hard and high/soft states were reported in IC 342 X-1 and X-2 with {\it ASCA} observations \citep{Kubota2001} and also in Holmberg IX X-1 with multiple missions \citep{LaParola2001}. However, almost no such canonical state transitions have been reported in the {\it Chandra} and {\it XMM-Newton} era. In contrast, transitions between soft and hard states were found but not in a way similar to Galactic sources \citep{Dewangan2004,Feng2006a,Feng2009,Dewangan2010}. A major difference between ULXs and Galactic BHBs is that most ULXs are persistent sources and have been bright since their discovery decades ago, while the latter are usually transients especially accreting from a low mass companion with Roche lobe overflow.

{\it Swift} monitoring \citep{Kaaret2009a,Strohmayer2009,Kong2010,Vierdayanti2010,Gris'e2010} of bright ULXs have revealed flaring activity with a luminosity change by a factor of $\lesssim 10$ on time scales of days to months. In particular, \citet{Kaaret2009a} found Holmberg IX X-1 remains in the hard state as its flux varies, and \citet{Gris'e2010} found the X-ray spectral state is not correlated with the luminosity in Holmberg II X-1. Also, as mentioned previously in Section~\ref{sec:curve}, \citet{Feng2010} found M82 X-1 may have changed from the hard state to thermal state during outbursts. 

There is one exceptional case. ESO243$-$49 HLX-1 shows a low/hard to high/soft jump reminiscent of the canonical state transition \citep{Godet2009,Servillat2011}. It also displays outbursts in a ``fast rise exponential decay'' pattern, very similar to those seen in Galactic BHBs \citep{Lasota2011}. However, the authors argued that the observed variability is unlikely triggered by disk instability due to inconsistent timescale.

\section{Fast X-ray variability and mass scaling}
\label{sec:var}

\subsection{Observational characteristics}

The power spectrum of Galactic BHBs can be decomposed into broad components (power-law red noise, band-limited noise, flat-top or zero-centered broad Lorentzian, etc.) and narrow components (QPOs). Although we still know little about the physical origin of these temporal features, their phenomenology has proved useful in defining and identifying accretion states \citep{Belloni2010} and possibly in determining BH masses via model independent calibrations \citep[\textit{e.g.}][]{McHardy2006}.
 
An important model-independent finding of \citet{Heil2009}, based on the study of sample of 16 bright ULXs, is that those sources can be divided into two groups defined by their short-term X-ray variability, even though they have similar energy spectra and luminosities. One group (including among others NGC 5204 X-1) shows weak or absent variability on timescales at least as large as 100s. The other group (including among others NGC 5408 X-1) has similar variability levels and power spectra as luminous Galactic BHBs and AGN in the observed frequency bandpass ($10^{-3}$--$1$ Hz). It is still unclear what suppresses short-term variability in one class but not the other. In Galactic BHBs, high variability is often associated to non-thermal emission with steady jets; low variability to thermal emission with no jets \citep{Belloni2010}. 

\begin{figure}[bt]
\centering
\includegraphics[width=\columnwidth]{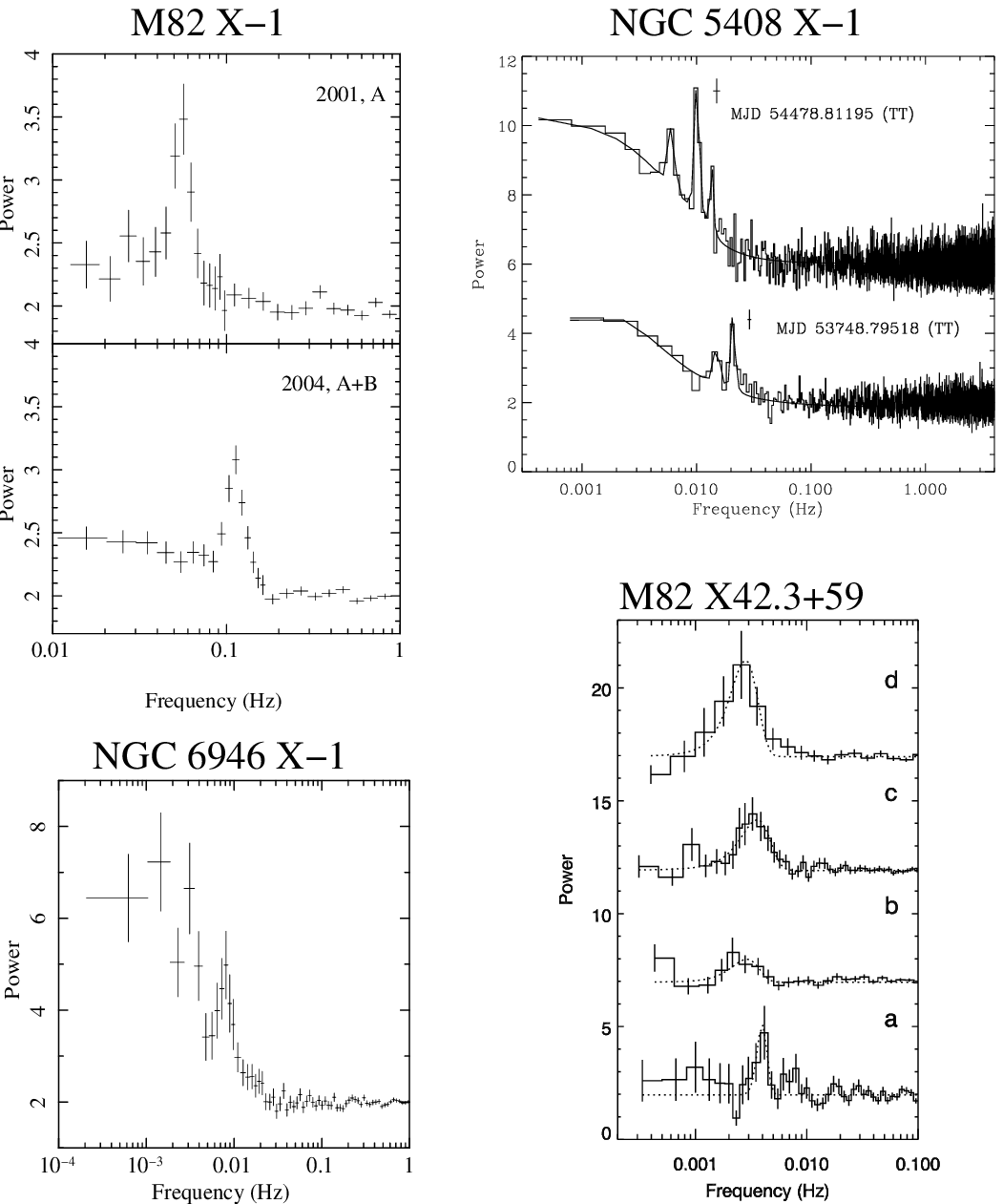}
\caption{Power spectra of four ULXs. M82 X-1: reproduced using data from \citet{Feng2007} with {\it XMM-Newton} observations in 2001 and 2004, respectively; A and B refer to the photon extraction regions. NGC 5408 X-1: the two spectra are respectively from {\it XMM-Newton} observations in 2006 and 2008 \citep[Figure~5 in][]{Strohmayer2009a}. NGC 6946 X-1: reproduced using data from \citet{Rao2010}. M82 X42.3+59: four spectra are from {\it Chandra} (b--d) and {\it XMM-Newton} (a) observations \citep[Figure 1 in][]{Feng2010a}.
\label{fig:psd}}
\end{figure}

In the following, we select four ULXs in which interesting structures have been found in their power spectra, shown in Figure~\ref{fig:psd}.

{\it M82 X-1, also known as M82 X41.4+60} --- \citet{Strohmayer2003} first discovered low frequency QPOs around 54~mHz from the M82 galaxy with both {\it XMM-Newton} and {\it RXTE} observations. By fitting the {\it XMM-Newton} images to its point spread function with known source positions from {\it Chandra}, \citet{Feng2007} identified that M82 X-1 is the source that produced the QPOs, and interestingly, it was not the brightest X-ray source in M82 at the time of QPO discovery. A following {\it XMM-Newton} observation revealed that the QPOs have changed the central frequency from 54~mHz to 114~mHz, above a flat-top component with a low frequency break at 34~mHz \citep{Dewangan2006,Mucciarelli2006}. \citet{Mucciarelli2006} also found that the QPOs changed frequency from 107~mHz to 120~mHz during the second observation, and a plausible harmonic ratio of 1:2:3 for QPO frequencies with {\it RXTE} data. These QPOs, however, were undetected with three {\it XMM-Newton} observations of similar depth, suggestive of a spectral state transition \citep{Feng2010}.

{\it NGC 5408 X-1} --- With a short {\it XMM-Newton} observation, \citet{Soria2004} found the source is highly variable, and its power spectrum has a break at 2.5~mHz. \citet{Strohmayer2007} reported the identification of narrow QPOs around 20~mHz from the source, and confirmed the presence of the frequency break. The QPOs are narrow and may show a pair with a 4:3 frequency ratio. A longer {\it XMM-Newton} observation later detected the QPOs at about 10~mHz and the energy spectral parameters of the source also changed \citep{Strohmayer2009a}. The QPO frequency was found to scale with the disk flux and also the spectral index of the power-law component. These correlations are analogous to that seen in Galactic BHBs, suggesting that this ULX is in an emission state similar to sub-Eddington sources. The QPO properties (narrow, strong, with a flat-top continuum) and the temporal-spectral evolution patterns suggest that the QPOs are of type C and the compact object in NGC 5408 X-1 is an IMBH. However, as the QPO frequency is not proportional to the break frequency of the continuum as expected, the identification as of type C was questioned and so was the IMBH scenario \citep{Middleton2011}; they proposed that the source is super-Eddington and the turbulence in the Comptonization region gives rise to the observed variability. Other authors have also rejected the IMBH argument \citep{Soria2007}, pointing out that neither lower QPO frequencies nor larger, cooler disks can be used as indicators of higher BH masses in ULXs; both findings may simply suggest that the visible inner radius of the disk has receded much further than the innermost stable orbit, and has been covered or replaced by a thick Comptonizing region or outflow in the inner region. Besides, \citet{Heil2010} also found that the source shows a linear correlation between the rms amplitude and flux, similar to that seen in bright BHBs and AGN.

{\it M82 X42.3+59} --- This is a transient ULX in M82 \citep{Matsumoto2001}. It could not be detected by {\it Chandra} in its low state, but sometimes, was even brighter than M82 X-1, being the most luminous source in M82 \citep{Feng2007}. With {\it Chandra} observations, and also confirmed by a simultaneous {\it XMM-Newton} observation once, \citet{Feng2010a} discovered low frequency QPOs around 3-4~mHz in its power spectrum. These QPOs are wide and strong, appear above white noise, and can only be detected when the source is brighter than $10^{40}$~\ergs. This is consistent with a Type A or B classification, but is inconsistent with the properties of type C QPOs. 

{\it NGC 6946 X-1} --- This source has the highest short-term variability found in a ULX, with a fractional rms amplitude of 60\% integrated in the frequency range of 1-100~mHz \citep{Rao2010}. Its power spectrum shows a flat-top continuum that breaks at about 3~mHz with possible quasi-periodic oscillations (QPOs) near 8.5~mHz.

\subsection{Implications on the mass}

The characteristic frequency alone usually is not a direct indicator of the BH mass, as these noise features can vary dramatically in a single source. However, correlations between the frequency and spectral parameters have been found, and the correlation patterns are thought to scale with the BH mass.

The most common QPOs seen in BHBs are of type C \citep[\textit{cf.}][]{Casella2005}. The frequency of type C QPOs are found to vary over a wide range. For example, the lowest detected frequencies in GRS 1915+105 are around 1~mHz \citep{Morgan1997}, even lower than the frequencies of all QPOs detected in ULXs. A positive correlation between the QPO frequency and disk flux has been found for type C QPOs in XTE~J1550$-$564, GRO~J1655$-$40 and H~1743$-$322 \citep{Sobczak2000,Remillard2006,McClintock2009}. The QPO frequency also scales with the photon index of the power-law component in the energy spectrum \citep[\textit{e.g.}][]{Sobczak2000,Vignarca2003}. \citet{Titarchuk2004} suggested that the QPO frequency scales inversely with the BH mass at a fixed power-law photon index, and thus one can infer the mass given spectral index \citep{Shaposhnikov2009}. On the contrary, Type A and B QPOs show in a relatively narrow frequency range. 

The broad continuum seems to be a more reliable determinant of BH mass, as it is detected in both BHBs and AGN and the scaling has been tested over a wide frequency and mass range. sMBHs and AGN appear to populate a ``variability plane'' defined by the BH mass, accretion rate, and $\nu_l$, which is the frequency of the lower high-frequency Lorentzian component in the power spectrum \citep{McHardy2006,Kording2007}. Limited by sensitivity, $\nu_l$ cannot be detected in ULXs. Based on a correlation between $\nu_l$ and the frequency of type C QPOs,  \citet{Casella2008} extended the variability plane to type C QPOs that are visible in ULXs.  The BH mass can be inferred from the X-ray luminosity (here we adopt the radiatively efficient accretion case) and type C QPO frequency as
\begin{equation}
\log (M/M_\odot) \gtrsim 0.5 \log(L_{\rm X}/0.1c^2) - 0.51 \log \nu_{\rm QPO} - 7.9 \; ,
\end{equation}
where the inequality is due to the unknown bolometric correction ($L_{\rm X} \lesssim L_{\rm bol}$). Although the ``variability plane'' is tested for a wide range of BH masses,  the correlation between $\nu_l$ and $\nu_{\rm QPO}$ has not been tested in ULXs and even the identification as of type C is questioned at least for NGC 5408 X-1 \citep{Middleton2011}.

Attempts to scale BH mass with the variability amplitude (rms) have made great success in AGN \citep{Zhou2010}. However, it is found that ULXs are significantly less variable than AGN, and a simple extrapolation could not be obtained \citep{Gonz'alez-Mart'in2011}.

Based on the correlations mentioned above, the BH masses inferred from the characteristic frequencies are all about $10^3$--$10^4$~$M_\odot$, indicating that ULXs contain IMBHs. However, application of these mass scaling methods requires strong caveats. There are at least three problems. First, we are not sure whether the timing features detected in ULXs are the same type found to scale with BH masses in Galactic BHBs; for example, we are not sure whether the QPOs in NGC 5408 X-1 are of Type C \citep{Middleton2011}. For other types of timing features (\textit{e.g.}, type A and B QPOs), it is still not known whether their frequency is proportional to the BH mass \citep{Casella2005}. Second, frequency-luminosity-mass relations have been calibrated and tested for Galactic BHBs and AGN in sub-Eddington accretion, that is when the inner edge of the disk is constant or moves towards the innermost stable orbit as the accretion rate increases (thus, increasing QPO frequencies). There is no empirical evidence that the scaling goes the same way in ULXs, which may be super-Eddington systems where the visible inner edge of the disk recedes from the innermost stable orbit for increasing accretion rates (thus, reducing QPO frequencies). The third problem is that we do not have enough signal-to-noise ratio to detect high-frequency QPOs, which severely limits possibility of direct comparisons with Galactic BHBs.

\section{Optical counterparts and stellar environment}
\label{sec:opt}

In this section the ``optical counterpart'' refers to the point-like optical source that is spatially associated with the ULX; sources with resolved structure, such as optical and radio nebulae, will be discussed in the next section. The optical emission from the counterpart could arise from the donor star or outer accretion disk, or both. It provides interesting information regarding the binary evolution history, nature of the donor star, disk geometry, and mode of mass transfer, and would also help constrain the BH mass.

Several groups have modeled X-ray irradiation in order to distinguish between the contributions of the outer disk and the donor star, and constrain the intrinsic stellar properties. \citet{Copperwheat2005,Copperwheat2007} applied their irradiation model to the photometric data for seven ULXs; they suggested that donor stars are consistent with spectral type B or later, and the BH masses are in the range of $\sim 10$--$100 M_{\odot}$. \citet{Patruno2008,Patruno2010} computed evolution tracks of ULX counterparts on the color-magnitude diagram, assuming a massive companion orbiting an sMBH or a MsBH; they applied their models to NGC\,1313 X-2, suggesting a BH mass $\sim 50$--$100 M_{\odot}$. In addition to reproducing the observed colors and luminosity (after accounting for irradiation and disk contribution), a donor star in a ULX must be able to supply a mass accretion rate $\gtrsim 10^{-6} M_{\odot}$ yr$^{-1}$, which requires Roche lobe overflow, and to do it over a sufficiently long period of time to explain the observed number of ULXs per galaxy. Binary population synthesis models have been developed and used to address this problem \citep{Podsiadlowski2003,Rappaport2005,Madhusudhan2006,Madhusudhan2008}. Modeling the binary evolution requires calculations of the mass transfer rate, the corresponding change in orbital separation, and the subsequent fate of the transferred mass. This in turn depends critically on whether we impose a strict Eddington limit on the mass transfer rate that can be accepted by the BH (with the rest being expelled in a wind), or instead we assume that the BH can accrete at higher rates. Thus, for the same kind of donor star, the outcome in terms of X-ray luminosity and lifetime is very different depending on whether the primary is an Eddington-limited stellar BH, a non-Eddington-limited stellar BH, or an IMBH. \citet{Madhusudhan2008} considered three characteristic families of binary evolution tracks, with different BH mass distributions: $6$--$15 M_{\odot}$, $6$--$24 M_{\odot}$, and a single-mass population of $1000 M_{\odot}$ BHs. They built probability diagrams in different slices of parameter space (optical color-magnitude diagram; orbital period versus potential X-ray luminosity; X-ray luminosity versus age), representing the total time spent by all the systems in those regions of parameter space. Their main conclusion is that the most probable ULX system parameters correspond to high-mass donors ($\gtrsim 25 M_{\odot}$), with orbital period between 1 and 10 days and characteristic age $\sim 10^7$ yr. In terms of BH mass, they found that both the $1000 M_{\odot}$ BH population, and that with non-Eddington-limited BHs up $25 M_{\odot}$ were consistent with the optical and X-ray observations and with the observed number of ULXs per galaxy; the $25$--$100 M_{\odot}$ BH mass range has not been tested yet.

Thanks to the sub-arcsecond resolution of {\it Chandra} and {\it HST}, the optical counterparts of more than a dozen ULXs have been identified. These are done mainly by aligning {\it Chandra} and {\it HST} images using objects with emission on both, to reduce the {\it Chandra} X-ray position error from 0.6\arcsec\ to about 0.2\arcsec--0.4\arcsec.  \citet{Tao2011} collected all available data in the {\it HST} archive to investigate the optical properties of 13 ULXs, which can be summarized as follows.

\begin{enumerate}[a)]

\item The X-ray to optical flux ratios and optical colors are consistent with that of LMXBs, but unlike HMXBs or AGN. This indicates that the optical emission from ULXs is mainly due to reprocessing of X-rays on the outer accretion disk. As a consequence, one cannot infer the spectral type of the companion star using optical colors directly.

\item The broad band optical spectrum can be generally fitted with a power-law model ($F_\nu \propto \nu^\alpha$). The distribution of the spectral index $\alpha$ peaks between 1 and 2, which is also consistent with disk irradiation models.

\item The counterpart has unique optical properties within its neighborhood. Nearby sources usually show a blackbody spectrum instead of a power-law spectrum.

\item The optical emitting region is estimated to be large, of the order of $10^{12}$~cm, indicative of long binary periods ($\gtrsim 10$~days) for sMBH or MsBH primaries, or shorter periods for IMBHs. 

\item Sources with multiple observations usually show strong, random variability on both magnitudes and colors.

\end{enumerate}

There are a few outliers showing peculiar properties. The optical spectra of NGC 2403 X-1 and M83 IXO 82 have a slope consistent with that of the standard disk emission, $F_\nu \propto \nu^{1/3}$. M101 ULX-1 and M81 ULS1 are more similar to HMXBs than LMXBs in terms of X-ray to optical flux ratios. IC 342 X-1 may show a blackbody-like spectrum instead of a power-law. More details can be found in \citet{Tao2011}.

There have been a lot of observational efforts to determine the nature of the companion stars in nearby ULXs.  So far, NGC 7793 P13 is the only ULX in which emission from the companion star is directly seen beyond reasonable doubt \citep{Motch2011}; detection of narrow emission lines identifies the companion star as a late-B type supergiant with a mass $\approx 20 M_\odot$. 

Orbital periods may have been detected in a few sources. Assuming Roche-lobe overflow, the companion density can be simply determined from the binary period, $\rho = 115 P_{\rm hr}^{-2}$~g~cm$^{-3}$ \citep{Frank2002}. M82 X-1 is the strongest case of a positive detection; a monitoring program with {\it RXTE} revealed a steady period of about 62~day \citep{Kaaret2006,Kaaret2006a,Kaaret2007}, corresponding to a companion density of $5 \times 10^{-5}$~g~cm$^{-3}$. \citet{Strohmayer2009} discovered a 115~day X-ray period from NGC 5408 X-1 with {\it Swift} monitoring, and inferred a similar companion density of $1.5 \times 10^{-5}$~g~cm$^{-3}$; however, the interpretation of that modulation as a binary period has been disputed \citep{Foster2010}. These claimed long periods imply an evolved donor star. Possible detections include a 6.1 day period from NGC 1313 X-2 with {\it HST} observations \citep{Liu2009}; however, low significance of the period was claimed by other groups \citep{Gris'e2008,Gris'e2009,Impiombato2011}. 

M82 X-1 is located near or within a super star-cluster MCG-11 with an age of 7-12 Myr \citep{McCrady2003}. NGC 1073 IXO 5 appears near a ring of star formation of 8--16 Myr old \citep{Kaaret2005}. NGC 1313 X-2 is found near a star cluster with an age of 20~Myr \citep{Liu2007,Gris'e2008}. IC 342 X-1 lies in a field with relatively old stellar objects, mostly older than 10~Myr \citep{Feng2008}. The stellar field around NGC 1313 X-1 is at least 30~Myr old  \citep{Yang2011}. Holmberg IX X-1 is suggested to be part of a loose cluster with an age $\lesssim 20$~Myr \citep{Gris'e2011}. All those findings suggest that ULXs are in relatively old stellar environments with ages of at least 10~Myr. 

An exceptional case is provided by a transient ULX recently discovered in M83 \citep{Soria2010}. It is located in an interarm region with a stellar population older than $\sim 10^8$ yr. {\it HST} observations when the system was probably in quiescence constrain the donor star to be $< 5 M_{\odot}$. Instead, {\it Gemini} observations during the recent outburst revealed a bright, blue counterpart with absolute $B$ magnitude $\sim -4$, due to the irradiated outer disk and/or irradiated donor star.  

Identification of the optical counterparts to ULXs is the first step for a potential dynamical mass measurement. \citet{Pakull2006} found that the broad He {\sc ii} $\lambda4686$ line from NGC 1313 X-2 changed its velocity by about 380~km~s$^{-1}$ between two observations. If this is due to binary motion of the disk, an upper limit on the BH mass of 50~$M_\odot$ can be placed. There have been a few attempts to obtain the mass function of a few ULXs using emission-line velocity curves. However, the lines were found to vary randomly instead of showing an ellipsoidal modulation; thus, they did not provide dynamical constraints on the BH masses \citep{Roberts2011}. The orbital period is a key parameter for dynamical mass measurement. With current facilities, it may be more feasible to search for orbital periods in X-rays and then confirm them in optical for extragalactic sources, like the cases of M33 X-7 and IC 10 X-1.

\section{Jets, outflows and bubbles}
\label{sec:bub}

\subsection{X-ray photoionized nebulae}

We naturally expect ULXs to ionize the surrounding interstellar medium \citep[see][]{Tarter1969,Kallman1982} for the basic principles of X-ray ionization. The main difference between UV ionized \ion{H}{ii} regions and X-ray ionized nebulae is the lack, in the latter, of a sharp boundary (Str\"omgren sphere) between the ionized and neutral plasma \citep{Pakull2002}. Instead, there is an extended warm region where weakly ionized atoms coexist with collisionally excited neutral species. As a result, an X-ray ionized nebula is characterized by high-ionization emission lines such as \ion{He}{ii} $\lambda$4686 and [\ion{O}{iv}] at 25.89 $\mu$m in the inner regions, and low-ionization forbidden lines such as [\ion{O}{i}] $\lambda$6300 in the outer regions. In particular, \ion{He}{ii} $\lambda$4686 acts as a photon counter of the ionizing flux between 54 eV and $\approx 200$ eV \citep{Pakull1986}. By comparing the true source flux inferred from the \ion{He}{ii} $\lambda$4686 line with the apparent flux measured at energies $\gtrsim 300$ eV with {\it Chandra} or {\it XMM-Newton}, we can test whether the X-ray emission is isotropic or beamed.

The N159F nebula, powered by the sMBH LMC X-1, is the best-known example of a large-scale ($\sim 10$ pc) X-ray ionized nebula in Galactic systems \citep{Pakull1986}. \citet{Pakull2002,Pakull2003} opened a new field with their systematic search for similar nebulae around ULXs. They found a spectacular case around the ULX in M81 group dwarf galaxy Holmberg II. The Holmberg II X-1 ionized nebula (now known as ``Foot Nebula'') has a \ion{He}{ii} $\lambda$4686 luminosity $\approx 2.5$--$2.7 \times 10^{36}$ erg s$^{-1}$ \citep{Pakull2002,Kaaret2004}, 30 times higher than the LMC X-1 nebula, which implies an average\footnote{over the recombination time of He$^{++}$ in the nebula, estimated as $\sim 3000$ yr \citep{Kaaret2004}.} ionizing X-ray flux $\sim 4$--$6 \times 10^{39}$ erg s$^{-1}$. This result has been confirmed by further optical/infrared spectroscopic studies \citep{Lehmann2005,Abolmasov2007,Berghea2010}. On the other hand, the apparent X-ray luminosity of Holmberg II X-1 has been seen to vary over the years between $\sim$ a few $10^{39}$ erg s$^{-1}$ and $\approx 3 \times 10^{40}$ erg s$^{-1}$ \citep{Zezas1999,Miyaji2001,Goad2006,Gris'e2010,Caballero-Garc'ia2010}. The consistency between the two independent measurements gives the strongest evidence to date that beaming is negligible ($1/b \lesssim 3$) for this ULX; this favors models with relatively large BH masses (Section \ref{sec:bh}). Similar arguments have also been applied to the photoionization-dominated nebula around NGC 5408 X-1. From the flux in high-ionization lines such as \ion{He}{ii} $\lambda$4686 and [\ion{Ne}{v}] $\lambda$3426, \citet{Kaaret2009} inferred a lower limit on the isotropic X-ray luminosity $\approx 3 \times 10^{39}$ erg s$^{-1}$, only a factor of 3 lower than the directly measured X-ray luminosity. As in Holmberg II X-1, this suggests that beaming factor is $\lesssim 3$. More examples of X-ray ionized nebulae in ULXs are reported by \citet{Pakull2002}, \citet{Abolmasov2007}, and \citet{Kaaret2010}, although in those cases the interpretation is more model dependent, because the observed line spectra are a mix of photo-ionized and shock-ionized plasma emission.

\subsection{Shock-ionized nebulae}

\begin{figure*}[htbp]
\centering
\includegraphics[height=0.4\textwidth]{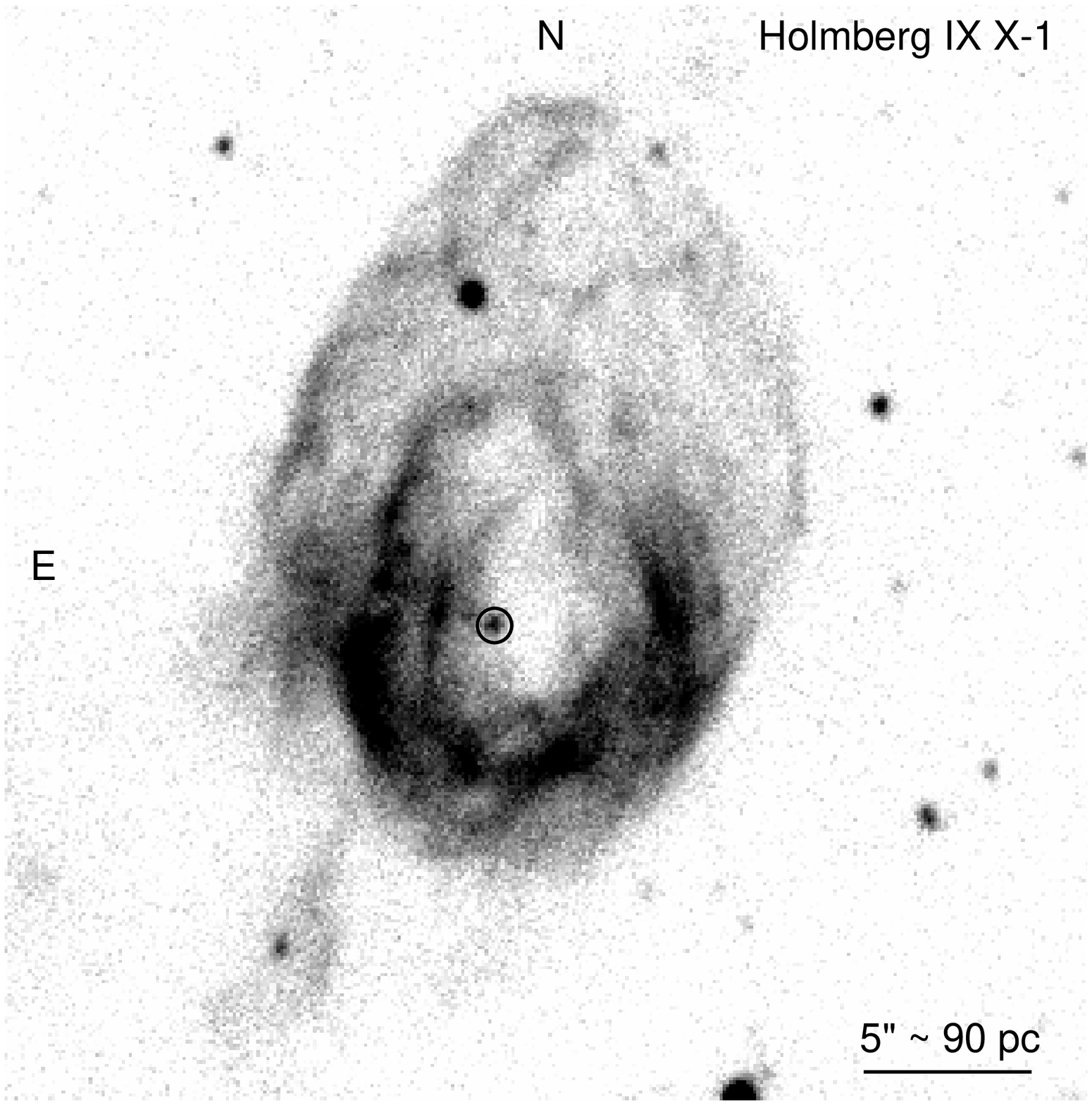}
\includegraphics[height=0.4\textwidth]{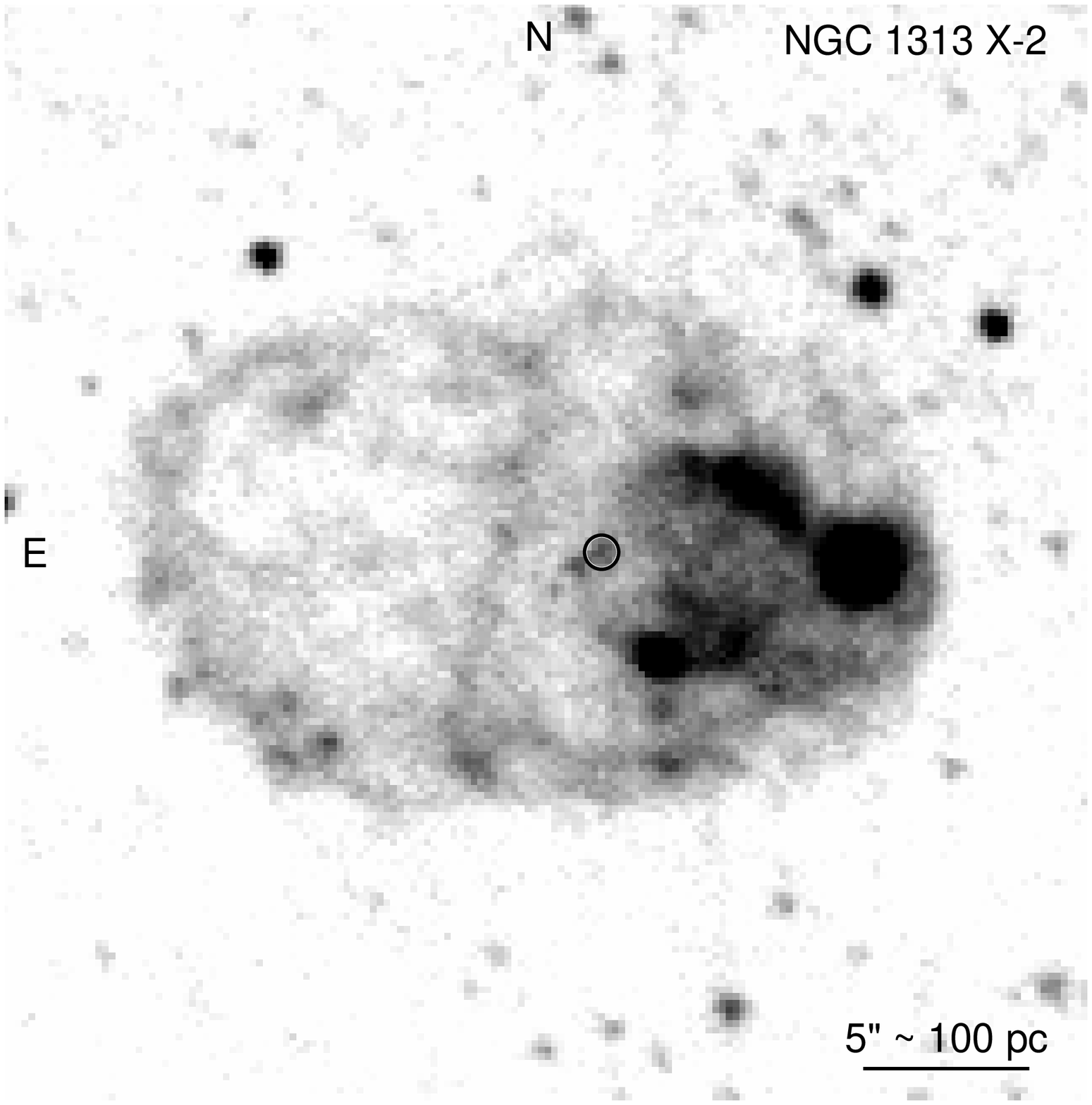}\\
\includegraphics[height=0.4\textwidth]{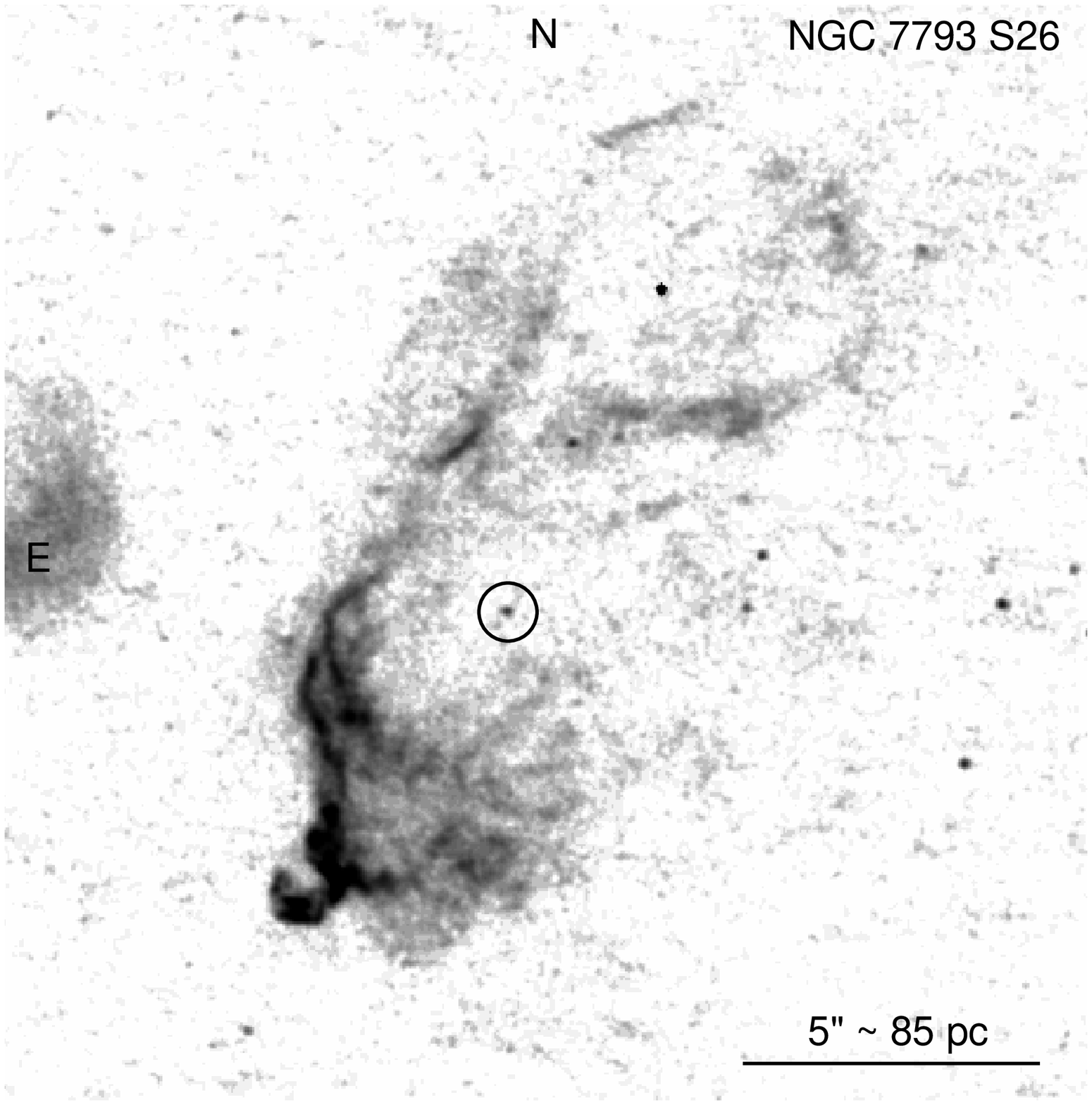}
\includegraphics[height=0.4\textwidth]{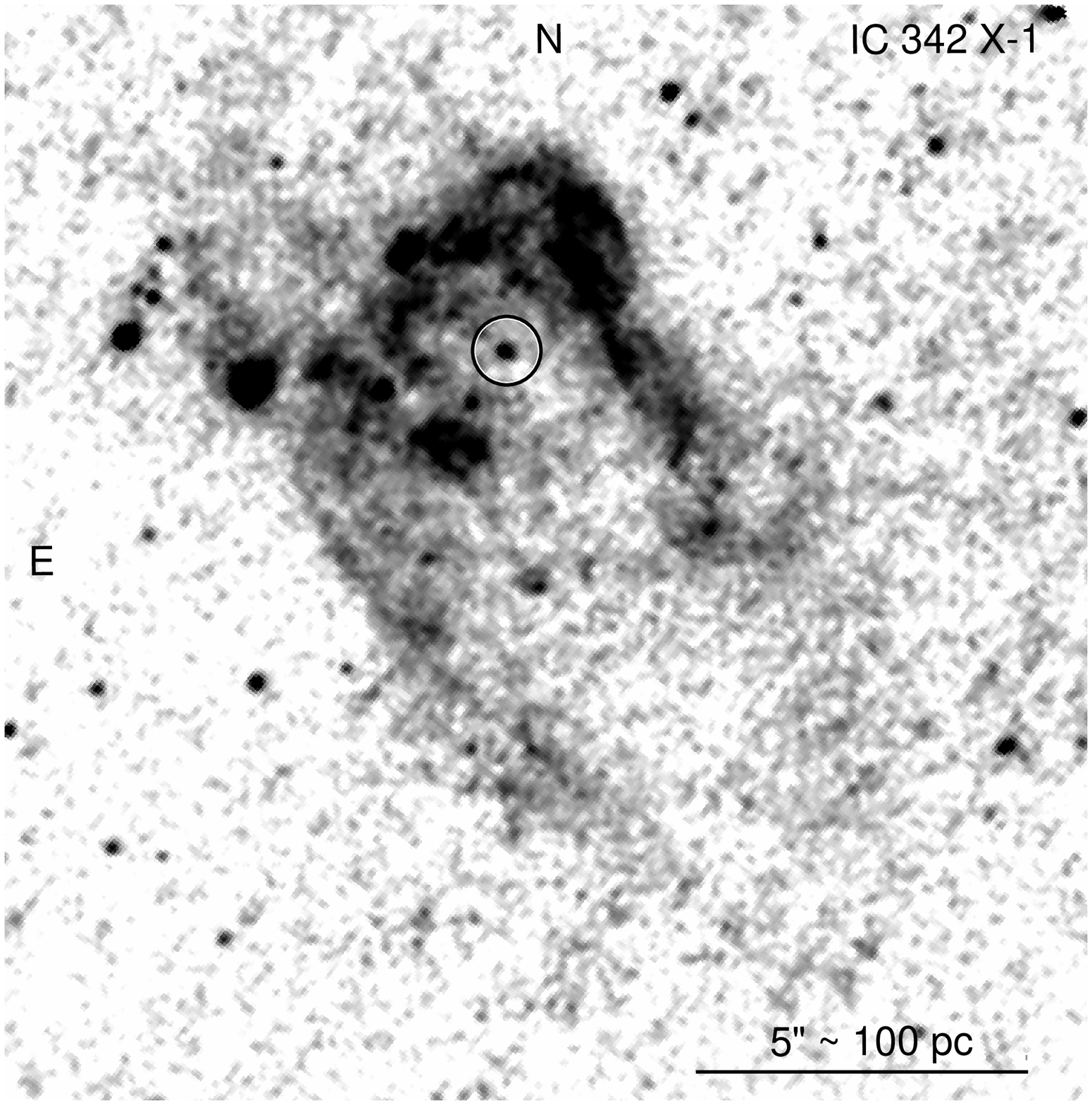}\\
\caption{Large H$\alpha$ emission nebulae aroud a selection of ULXs and in the jet-dominated source NGC\,7793 S26. Clockwise, from top left: Holmberg\,IX X-1 ({\it Subaru}/FOCAS image from 2003, credit F. Gris\'{e}); NGC\,1313 X-2 ({\it VLT}/FORS1 image from 2003, credit F. Gris\'{e}); IC\,342 X-1 ({\it{HST}}/ACS image from 2005); NGC\,7793 S26 ({\it{HST}}/WFC3 image from 2011). In each panel, the X-ray source position is marked with a circle of 0.5\arcsec\ radius. 
\label{fig:bubble}}
\end{figure*}

It is often hard to distinguish between X-ray photoionization and shock excitation, for ULX nebulae---just as it is generally difficult to separate the contribution of a starburst nucleus and/or a low-luminosity AGN in a LINER. Diagnostic diagrams with several line ratios are computed with shock and photoionization codes such as {\sc mappings iii} \citep{Allen2008} and {\sc cloudy} \citep{Ferland1998}. Shock-ionized plasma is characterized by high ratios of low-excitation lines such as [\ion{O}{i}] $\lambda 6300$ [\ion{S}{ii}] $\lambda\lambda 6717$,$6731$ and [\ion{N}{ii}] $\lambda\lambda 6548$,$6583$ over H$\alpha$. For shock velocities $v_{\rm s} \gtrsim 80$--$100$ km s$^{-1}$, the high-ionization, forbidden transitions [\ion{O}{iii}] $\lambda\lambda 4959$,$5007$ are strong ([\ion{O}{iii}] $\lambda$ 5007/H$\alpha \approx 1$ for $v_{\rm s} \gtrsim 100$ km s$^{-1}$), while \ion{He}{ii} $\lambda$4686 indicates $v_{\rm s} \gtrsim 300$ km s$^{-1}$ \citep{Dopita1984}. The luminosity in the H$\beta$ line is only a function of the total shock power $P_{\rm w}$ and shock velocity \citep{Dopita1995}; for typical speeds $\sim 100$--$300$ km s$^{-1}$, $P_{\rm w} \approx 100$--$300 L_{{\rm H}\beta}$.

Supershells were first found around NGC 6946 MF 16 \citep{Blair1994} and Holmberg IX X-1 \citep{Miller1995}, and were initially speculated to be powered by unusually energetic SN remnants. The X-ray emission could not be from the nebulae, because of its variability \citep{Roberts2003}. Following the pioneering work of \citet{Pakull2002,Pakull2003}, several more huge shock-ionized nebulae have been identified around ULXs (``ULX bubbles'', see Figure~\ref{fig:bubble} for examples), with characteristic diameter $\sim 200$--$400$ pc, expansion velocity $\sim 100$--$200$ km s$^{-1}$, characteristic age $\sim 0.5$--$1$ Myr, mechanical power $\sim 10^{39}$--$10^{40}$ erg s$^{-1}$ (independently determined from the bubble expansion speed and from the H$\beta$ luminosity; \citealt{Weaver1977,Dopita1995}) and energy content $\sim 10^{53}$ erg \citep{Roberts2003a,Gris'e2006,Ramsey2006,Abolmasov2007,Pakull2008,Kaaret2010,Moon2011,Russell2011}. The two best-studied examples are the bubbles around NGC 1313 X-2 and Holmberg IX X-1 \citep{Pakull2002,Pakull2008}. ULX bubbles resemble ordinary SN remnants, but are an order of magnitude larger and 10 to 100 times more luminous. They cannot come from (hypothetical) single hypernova explosions: first, because the surrounding stellar populations are already too old (typically $\gtrsim 10^7$ yr) and without O stars; secondly, because it is unlikely that a lower-mass binary companion that survived the hypothetical hypernova explosion could evolve to fill its Roche lobe and feed the ULX already within $\lesssim 1$ Myr \citep{Pakull2008}. Invoking 10 to 100 ordinary SN explosions over the last 1 Myr would require a massive star cluster with $M \gtrsim$ a few $10^5 M_{\odot}$ ({\it e.g.}, from {\sc STARBURST99}; \citealt{Leitherer2010,Leitherer1999}). However, none of those ULXs is associated with a massive star cluster: at most, they are inside ordinary OB associations with age $\sim 10$--$20$ Myr and total stellar mass $\sim 10^4 M_{\odot}$. The only plausible conclusion is that ULX bubbles are inflated by a continuous jet or outflow from the ULX, with a power comparable to the radiative output \citep{Pakull2006,Pakull2008}.

Jet-inflated bubbles around accreting BHs are found in our Galaxy, too. The best-known example is the W50 nebula around SS\,433 \citep{Dubner1998,Goodall2011}: a spherical shell with two lateral extensions, the ears, that are inflated by interaction of the jets with the interstellar medium. The characteristic diameter along the major axis is $\approx 100$ pc. Both the jet power and the radiative luminosity are $\gtrsim 10^{39}$ erg s$^{-1}$. Indeed, it is suggested that SS\,433 would appear as a beamed, super-Eddington ULX if it was seen face-on \citep{Fabrika2004}. Another local example is the 5-pc bubble inflated by the sMBH Cygnus X-1 \citep{Gallo2005,Russell2007}: the shock velocity determined from optical emission line ratios is $\sim 100$--$360$ km s$^{-1}$; the required jet power $\approx 10^{37}$ erg s$^{-1}$ is $\sim 0.3$--$1$ times the X-ray luminosity.

\subsection{Jet-dominated sources}

The ULX bubbles discovered so far have a quasi-spherical appearance, with no direct signature of relativistic, collimated jets (as in the Galactic source SS\,433). The first extra-Galactic super-bubble with collimated jets was recently found in the outskirts of the spiral galaxy NGC\,7793 \citep{Pakull2010}. This source (Figure~\ref{fig:bubble}), known as S26, has: X-ray and optical hot spots where the jet and counter-jet interact with the interstellar medium; an optical nebula with a size of $\approx 300 \times 150$ pc, expanding at $\approx 250$ km s$^{-1}$; an associated X-ray and radio nebula with bright radio lobes. The long-term-average jet power $\sim$ a few $10^{40}$ erg s$^{-1}$ \citep{Pakull2010,Soria2010a}, over a characteristic age $\approx 2 \times 10^5$ yrs. The X-ray luminosity of the central BH is only $7 \times 10^{36}$ erg s$^{-1}$. This might be because the BH emission is highly obscured and we only see a scattered component; or because it is beamed away from us; or because the BH is a transient, currently in a radiative-faint state, but may appear as a ULX at other times. Whatever the reason for the current radiative dimness, it is important to note that NGC\,7793 S26 has had collimated jets with a long-term average power at least of order of the Eddington power (and probably much greater) of a stellar BH (sMBH or MsBH). In other words, it is a jet-dominated system at super-Eddington mass accretion rates. There is no physical principle that prevents mechanical power to exceed the Eddington luminosity. However, the fact that it is collimated in a large-scale jet may be surprising. The canonical view \citep{Fender2004} of BH accretion states---based mostly on the behavior of Galactic transients---is that collimated, steady jets are found at $\dot{m} \lesssim 0.03$, when accretion is radiatively inefficient (low/hard state). Whether and how steady, collimated jets can be launched during super-Eddington accretion phases remains an unsolved theoretical problem. A possible solution comes from the radiation-magnetohydrodynamic simulations of \citet{Takeuchi2010}, who find that a jet can be accelerated up to mildly relativistic speed by radiation pressure, and collimated by the Lorentz force of a magnetic tower made of wound-up toroidal magnetic field lines. The magnetic structure is itself kept collimated by the geometrically thick accretion flow that dominates in the innermost region when $\dot{m} \gtrsim 1$.

An ultrapowerful (rather than ultraluminous) system such as NGC\,7793 S26 is the local-universe analog of a recently discovered class of quasars, dominated by mechanical power \citep{Punsly2007,Punsly2011}. In those quasars, the jet power is up to 25 times higher than the bolometric radiative luminosity; in some of those systems, the jet power is near or above the Eddington power. Mechanically-dominated, radiatively inefficient quasars in the early universe may grow much faster than Eddington-limited sources, and would provide strong mechanical rather than radiative feedback onto the surrounding gas. Thus, understanding the inflow/outflow/radiation processes in ULXs has wider astrophysical relevance. Moreover, radiative and mechanical feedback from HMXBs and ULXs themselves may have been an additional,
important source of heating and reionization of the intergalactic medium at high redshifts \citep{Mirabel2011}.

\subsection{Radio bubbles}
\label{sec:radio}

Some ULX bubbles (from both the photoionized and shock-ionized variety) are associated to extended radio emission. The best-studied examples are Holmberg II X-1 \citep{Miller2005} and NGC\,5408 X-1 \citep{Kaaret2003,Soria2006,Lang2007}. Both radio nebulae are dominated by optically thin synchrotron emission, somewhat similar but much larger ($\approx 60 \times 40$ pc and $\approx 40$ pc, respectively) and more luminous than typical radio SN remnants. They are also an order of magnitude more luminous than the radio nebula around SS\,433 \citep{Dubner1998}, which well matches the relative values of radiative and mechanical core power in those sources. The total power carried by synchrotron-emitting electrons has been estimated to be between $\sim$ a few $10^{49}$ erg s$^{-1}$ and $\approx 10^{51}$ erg s$^{-1}$, depending on the very uncertain lower-energy cut-off of the distribution. In any case, both nebulae are much more energetic than W50 around SS\,433 ($\sim$ a few $10^{46}$ erg s$^{-1}$: \citealt{Dubner1998}) or than the extended lobes of typical Galactic microquasars such as GRS\,1758$-$258 ($\approx 2 \times 10^{45}$ erg s$^{-1}$; \citealt{Hardcastle2005}). NGC\,7793 S26 also has a large radio nebula \citep{Soria2010a}, interpreted as a combination of flat-spectrum free-free emission (mostly in the central region) and steep-spectrum synchrotron emission from the radio hot spots and lobes. The energy stored in relativistic electrons is $\sim$ a few $10^{50}$ erg s$^{-1}$.

The energy content in relativistic electrons for typical ULX nebulae is only $\sim 0.01$--$0.001$ of the core power---in particular, of the mechanical power when it can be reliably estimated from bubble size or optical line fluxes---integrated over the characteristic age of the bubble. The main reason is that most of the input energy is transferred to non-relativistic electrons, protons and nuclei \citep{Pakull2010,Soria2010a}. The fraction of energy in non-radiating particles to that in relativistic electrons in a system such as NGC\,7793 S26 is similar to the values found by \citep{Cavagnolo2010} for a sample of AGN, Seyfert and giant elliptical galaxies. For the same jet power, AGN, ULXs and Galactic microquasars may be radio louder or radio quieter depending on the amount of plasma entrained by the jets. We speculate that an interesting test may come from comparisons of the radio loudness and lobe energy in low/hard state microquasars, probably dominated by magnetically accelerated jets \citep{Komissarov2007}, and in ULXs, possibly dominated by radiation-pressure accelerated jets/outflows \citep{Takeuchi2010}, with more entrainment.

\section{Summary and future prospects}
\label{sec:sum}

There is no doubt that our understanding of the phenomenology of ULXs has greatly advanced in the past decade thanks to {\it Chandra}, {\it XMM-Newton}, and many other space- and ground-based facilities. However, the key physical issue is still unresolved: that is, whether they are powered by IMBHs or normal stellar BHs. A few sources have emerged as strong IMBH candidates: in particular, ESO243$-$49 HLX-1 and M82 X-1. On the other hand, for the majority of ULXs, there is no strong theoretical need nor compelling observational evidence for IMBHs. In the absence of direct mass-function measurements from phase-resolved optical spectroscopy, we still have to rely on X-ray spectral and timing modeling and other indirect clues. Reviewing the results that we have collected here, we suggest the following conservative statement:

\begin{quote}
{\it ULXs are a diverse population; MsBHs with moderate super-Eddington accretion seem to be the easiest solution to account for most sources up to luminosities $\sim$ a few $10^{40}$ erg s$^{-1}$; strong beaming ($1/b \gtrsim 10$) can be ruled out for the majority of ULXs; IMBHs are preferred in a few exceptional cases.}
\end{quote}

Some of the following kinds of observations may provide further breakthroughs:
\begin{enumerate}[a)]

\item measurements of the fast timing behavior down to a time scale of sub-seconds may probe the inner part of the inflow: ideally, we would search for possible high frequency features (breaks and QPOs) that are found in Galactic BHs at frequencies $\sim 10^2$ Hz. This would require a new generation of X-ray telescopes with large effective area.

\item determining the relative contribution of thermal emission and Comptonization component is a key test for competing accretion models. For that, multilayer-coated X-ray telescopes with good sensitivity up to a few tens of keV are needed.

\item detection of, or a more solid upper limit to broad iron K lines may constrain the properties of the inner disk.

\item High-resolution X-ray spectroscopy with gratings or microcalorimeters may constrain the metal abundance and ionization state of the inflow, the mass and energy of the outflow,
and probe the absorption from the surrounding interstellar medium.

\item the ULX sample within $\sim 15$ Mpc is almost complete, and it does not contain enough sources to permit statistical studies of the luminosity distribution above $\approx$
a few $\times 10^{40}$ erg s$^{-1}$. A homogeneous all-sky survey, complete for ULXs up to $\sim 50$ Mpc would allow us to determine whether or not the luminosity distribution extends unbroken up to at least $10^{41}$ erg s$^{-1}$. Such survey will hopefully be successfully conducted by {\it eROSITA};

\item repeated snapshot monitoring of nearby ULXs over the next few years will extend their long-term lightcurves enough to prove that their duty cycles and transient behavior are different from those of Galactic BHs. This can be easily done with smaller X-ray missions;

\item phase-resolved spectroscopy would lead to the measurement of radial velocity curves and mass functions. Attempts to do so with 8-m class telescopes have so far been inconclusive but have suggested that it will be easily done with future 30-m class telescopes;

\item integral-field spectroscopic studies of star forming galaxies, as well as low-frequency radio studies, may reveal fainter, older ULX bubbles or lobes, still expanding into the interstellar medium but no longer energized by a central source. This would constrain the duration of the ULX active phase;

\item searching for compact radio jets down to a sensitivity of $\sim 1 \mu$Jy will constrain the BH mass, in relation to other classes of accreting BHs in the fundamental plane. This may be done in the near future with the {\it eVLA} and with {\it SKA} Pathfinders, and in the more distant future with the SKA itself.

\end{enumerate}

In conclusion, we gather from this list that there is more than one way to skin a cat. Many fundamental aspects of ULX physics will be understood with instruments already planned or under construction. Thus, we anticipate a fruitful decade of ULX studies beyond the {\it Chandra} and {\it XMM-Newton} era.

\section*{Acknowledgements}

We thank Andy Fabian, Sean Farrell, Luigi Foschini, Fabien Gris\'e, Phil Kaaret, Kip Kuntz, Manfred Pakull, Tim Roberts, Mat Servillat, Tod Strohmayer, Doug Swartz, Dany Vanbeveren, and Luca Zampieri for helpful discussions and comments. HF acknowledges funding support from the National Natural Science Foundation of China under grant No.\ 10903004 and 10978001, the 973 Program of China under grant 2009CB824800, the Foundation for the Author of National Excellent Doctoral Dissertation of China under grant 200935, the Tsinghua University Initiative Scientific Research Program, and the Program for New Century Excellent Talents in University. RS acknowledges support from a Curtin University senior research fellowship, and hospitality at the Mullard Space Science Laboratory, Tsinghua University and the University of Sydney during part of this work.


\end{document}